\documentclass[11pt]{article}

\usepackage[colorlinks,bookmarksopen,bookmarksnumbered,citecolor=red,urlcolor=red]{hyperref}
\usepackage{multirow}
\usepackage{url}
\usepackage{listings}
\usepackage{mathtools}
\usepackage{amsmath}
\usepackage{amsfonts}
\usepackage{footnote}
\usepackage{threeparttable}
\usepackage{enumitem}
\usepackage{caption}
\usepackage{breakurl}
\usepackage{booktabs}
\usepackage{color}
\usepackage{multicol}
\usepackage{authblk}
\makesavenoteenv{tabular}

\makeatletter
\renewcommand{\@seccntformat}[1]{}
\makeatother

\definecolor{mGreen}{rgb}{0,0.6,0}
\definecolor{mGray}{rgb}{0.5,0.5,0.5}
\definecolor{mPurple}{rgb}{0.58,0,0.82}
\definecolor{red}{rgb}{1, 0, 0}
\definecolor{backgroundColour}{rgb}{0.95,0.95,0.92}
\definecolor{darkGreen}{rgb}{0,0.8,0}

\lstdefinestyle{CStyle}{
    backgroundcolor=\color{backgroundColour},   
    commentstyle=\color{mGreen},
    keywordstyle=\color{magenta},
    numberstyle=\tiny\color{mGray},
    stringstyle=\color{mPurple},
    basicstyle=\footnotesize,
    breakatwhitespace=false,         
    breaklines=true,                 
    captionpos=b,                    
    keepspaces=true,                 
    numbers=left,                    
    numbersep=5pt,                  
    showspaces=false,                
    showstringspaces=false,
    showtabs=false,                  
    tabsize=2,
    language=C
}

\newcommand\BibTeX{{\rmfamily B\kern-.05em \textsc{i\kern-.025em b}\kern-.08em
T\kern-.1667em\lower.7ex\hbox{E}\kern-.125emX}}

\begin{document}


\title{Recovering single precision accuracy from Tensor Cores while surpassing\\ the FP32 theoretical peak performance}

\author[1]{Hiroyuki Ootomo}
\author[2]{Rio Yokota}
\affil[1]{School of Computing, Tokyo Institute of Technology}
\affil[ ]{\textit {ootomo.h@rio.gsic.titech.ac.jp}}
\affil[2]{Global Scientific Information and Computing Center, Tokyo Institute of Technology}
\affil[ ]{\textit {rioyokota@gsic.titech.ac.jp}}
\date{}

%
%

\newtheorem{assumption}{Assumption}

\maketitle

\begin{abstract}
Tensor Core is a mixed-precision matrix-matrix multiplication unit on NVIDIA GPUs with a theoretical peak performance of more than 300 TFlop/s on Ampere architectures.
Tensor Cores were developed in response to the high demand of dense matrix multiplication from machine learning. However, many applications in scientific computing such as preconditioners for iterative solvers and low-precision Fourier transforms can exploit these Tensor Cores.
To compute a matrix multiplication on Tensor Cores, we need to convert input matrices to half-precision, which results in loss of accuracy.
To avoid this, we can keep the mantissa loss in the conversion using additional half-precision variables and use them for correcting the accuracy of matrix-matrix multiplication.
Even with this correction, the use of Tensor Cores yields higher throughput compared to FP32 SIMT Cores. Nevertheless, the correcting capability of this method alone is limited, and the resulting accuracy cannot match that of a matrix multiplication on FP32 SIMT Cores.
We address this problem and develop a high accuracy, high performance, and low power consumption matrix-matrix multiplication implementation using Tensor Cores, which exactly matches the accuracy of FP32 SIMT Cores while achieving superior throughput. The implementation is based on NVIDIA's CUTLASS.
We found that the key to achieving this accuracy is how to deal with the rounding inside Tensor Cores and underflow probability during the correction computation.
Our implementation achieves 51TFlop/s for a limited exponent range using FP16 Tensor Cores and 33TFlop/s for full exponent range of FP32 using TF32 Tensor Cores on NVIDIA A100 GPUs, which outperforms the theoretical FP32 SIMT Core peak performance of 19.5TFlop/s.

\end{abstract}


\maketitle

\section{Introduction}

In order to meet the increasing demand of dense matrix-matrix multiplication from the machine learning community, processors with specialized computing units for matrix multiplication are being developed by numerous vendors. For instance, Google Tensor Processing Unit (TPU) \cite{jouppi_-datacenter_2017}, Intel Ponte Vecchio \cite{intel_corporation_ponte_2021}, IBM POWER10 \cite{ibm_corporation_ibm_2020}, Preferred Networks MN-Core \cite{preferred_networks_inc_mn-core_nodate} and NVIDIA GPUs, all have special arithmetic units for low-precision matrix-matrix multiplication.
The NVIDIA Tensor Core is a mixed-precision matrix-matrix multiplication unit on NVIDIA GPUs and its theoretical peak performance is more than 300 TFlop/s on the latest Ampere architecture \cite{nvidia_corporation_nvidia_nodate-1}.
Tensor Cores compute a matrix-matrix multiplication of two FP16 (IEEE 754 binary16) matrices in full-precision and accumulate in FP32 (IEEE 754 binary32). This results in higher accuracy in matrix-matrix multiplication compared to FP16 computing units.
This capability to perform fast matrix multiplication can be used not only by machine learning applications, but also scientific computing applications and middleware that support both communities.
Haidar \cite{haidar_harnessing_2018} uses Tensor Cores within a mixed-precision iterative refinement solver in order to exploit the speed of Tensor Cores while recovering the accuracy through refinement.
This method can be applied to recover full FP64 (IEEE binary64) accuracy and is currently used in MAGMA\footnote{\url{https://developer.nvidia.com/magma}} and NVIDIA's cuSOLVER implementation.\footnote{ {\tt cusolverIRSRefinement\_t} section of cuSOLVER Documentation  \url{https://docs.nvidia.com/cuda/cusolver/}}
Tensor Cores can also be used for sparse matrix multiplication in graph analytics, breadth-first search, multigrid methods, etc \cite{zachariadis_accelerating_2020}.
Furthermore, Tensor Cores have also been used for reduction/scan operations in Monte Carlo methods, sort algorithms, etc \cite{carrasco_analyzing_2018,dakkak_accelerating_2019,firoz_feasibility_2020}.

There have been several efforts to analyze the internal behavior of Tensor Cores.
Jia \textit{et al.} and Raihan \textit{et al.} analyze how Tensor Core assembly instructions divide the input matrices, and the order they compute multiplications of the subdivided matrices \cite{jia_dissecting_2018,raihan_modeling_2019}.
There have also been studies on how Tensor Cores support subnormal numbers and use RZ (Round toward Zero) \cite{fasi_numerical_2020}.
Others have performed error analysis of Tensor Cores, where the theoretical error bound of mixed-precision block FMA computation is analyzed and compared to the actual error of Tensor Cores \cite{blanchard_mixed_2020}.
Studies on error correction have also been proposed. We have mentioned earlier that the conversion of input matrices to FP16 results in a loss of accuracy.
To address this problem, Mukunoki \textit{et al.} uses the Ozaki scheme \cite{ozaki_error-free_2012} on Tensor Cores \cite{mukunoki_dgemm_2020}.
Using this method, it is possible to achieve single precision or even double precision accuracy on Tensor Cores.
This method is able to perform matrix-matrix multiplication in FP64 faster than the cuBLAS DGEMM on GPUs with limited FP64 support such as the NVIDIA GeForce series.
However, this method is slower when it comes to FP32, and is much slower than the cuBLAS SGEMM on \textit{any} GPU.
This method is also not competitive for FP64 matrix multiplication when compared to cuBLAS DGEMM on NVIDIA Tesla series GPUs.

For single-precision matrix-matrix multiplication, Markidis \textit{et al.} propose a method to improve the accuracy of Tensor Core computation by using auxiliary FP16 variables to account for the truncated bits \cite{markidis_nvidia_2018}.
Markidis' method and its extensions are used for FFT \cite{sorna_optimizing_2018}, QR Factorization \cite{ootomo_randomized_2020}, and quantum-based molecular dynamics simulations \cite{finkelstein_quantum-based_2021}.
However, the use of auxiliary FP16 variables alone is not sufficient to fully recover the FP32 accuracy.
Feng \textit{et al.} propose an improvement to Markidis' method, which uses a better rounding mode during the truncation to FP16 \cite{feng_egemm-tc_2021}.
However, they are still not able to match the accuracy of SGEMM with their error correction method.
We consider that there might be some technical errors in their paper.
First, they do not take into account the implicit bit in IEEE 754 floating-point numbers.
For example, they claim that FP16 has 10 mantissa bits so two FP16 numbers have a total of 20 mantissa bits.
However, with the implicit bit the total is actually $(10 + 1) \times 2 - 1 = 21$ bits.
This inaccurate description also causes some confusion during their description of the rounding they propose.
Second, their method truncates a single-precision value $x$ to a half-precision $x_{hi}$ and stores the remaining value $x - x_{hi}$ to $x_{lo}$.
They propose to decide the rounding of $x_{hi}$ by looking at the 21st bit of mantissa of $x$, but if we consider the implicit bit, they should be looking at the 22nd bit.
Furthermore, $x_{hi}$ will always store the first 10 bits when truncating, but $x_{lo}$ does not always store the next 10 bits.
This means that always looking at the same bit to decide the rounding of $x_{hi}$ will not result in the round-split method they intend to perform.
Therefore, their mantissa length analysis applied on Markidis' method (Truncate-Split) and their method (Round-Split; EGEMM-TC) might be incorrect.
In the end, Feng \textit{et al.} are not able to achieve an accuracy that exactly matches SGEMM \cite{feng_egemm-tc_2021}.

Another important advantage of matrix-matrix multiplication unit is energy consumption.
For instance, the top supercomputer listed (June 2021) in the Top500 --Fugaku--, requires 30MW of power to achieve 442 PFlop/s FP64 performance, and Exascale systems are predicted to consume even more power.
If we look at the Green500 list, the top systems are equipped with matrix-matrix multiplication hardware such as MN-Core, and NVIDIA A100. This reflects the energy efficiency of matrix-matrix multiplication hardware.
The power consumption of GPUs have been analyzed at the Parallel Thread Execution (PTX) level, middleware level, and application level. Sakamoto \textit{et al.} measured the effect of low-precision computing on power consumption of the ICCG method in their earthquake simulation \cite{sakamoto_effectiveness_2020}. Guo \textit{et al.} performed a CUDA PTX instruction level power analysis \cite{guo_verifying_2021}.

We envision that in the future a majority of applications will adopt mixed precision.
In this scenario, there will be variables that will be computed in FP64, FP32, and FP16.
For instance, the work by \cite{carson_accelerating_2018} uses three precisions during the iterative refinement and the LU is done in FP32, which calls an single-precision matrix-matrix multiplication.
Furthermore, while the current quantum computer simulation using tensor network contraction uses single-precision matrix-matrix multiplication, it has been also investigated that which part of computing precision is sensitive to the result \cite{liu_closing_2021}.
In recent year, the real machines of quantum computer have been developed and tried to be shown quantum supremacy, that they compute certain tasks that (classical) supercomputers are not be able to compute in realistic time.
Moreover, since they have low power consumption \cite{britt_quantum_2017}, energy efficiency is becoming an important metric when evaluating quantum supremacy.
For instance, qFlex is a quantum computer simulator based on tensor network contraction using single-precision complex matrix-matrix multiplication, where the power consumption of each component was reported during its simulation on Summit V100 GPUs \cite{villalonga_establishing_2020-1}.
Although they have considered to use FP16 and Tensor Cores in their simulation, they decided not to use it since FP16 has less exponent than FP32 and insufficient to use.

\begin{figure}[t]
    \centering
    \includegraphics[width=0.6\linewidth]{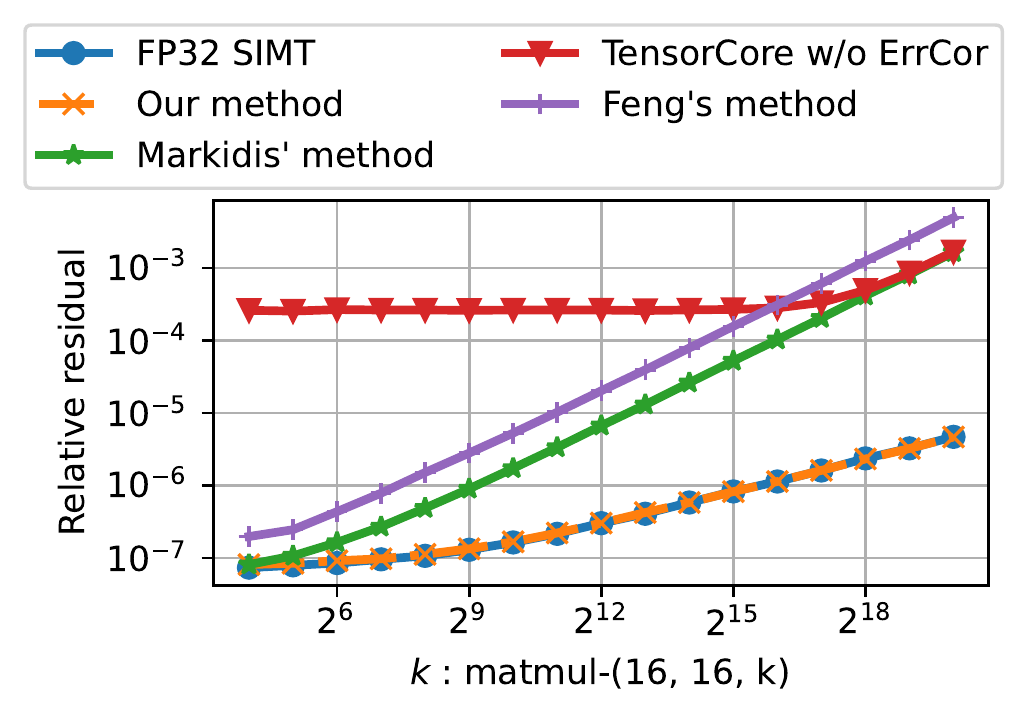}
    \caption{Accuracy comparison of matrix multiplication $\mathbf{A} \times \mathbf {B}$ of our method, Feng's method\cite{feng_egemm-tc_2021}, Markidis' method\cite{markidis_nvidia_2018}, cuBLAS SGEMM using FP32 SIMT Cores, and cuBLAS SGEMM using Tensor Cores. Input matrices $\mathbf{A} \in \text{FP32}^{16 \times k}$ and $\mathbf{B} \in \text{FP32}^{k \times 16}$ are initialized with random numbers generated from a uniform distribution $(-1, 1)$. The error is calculated by Eq. (\ref{eq:eval-matmul}).}
    \label{fig:intro-residual}
\end{figure}

\begin{figure}[t]
    \centering
    \includegraphics[width=0.6\linewidth]{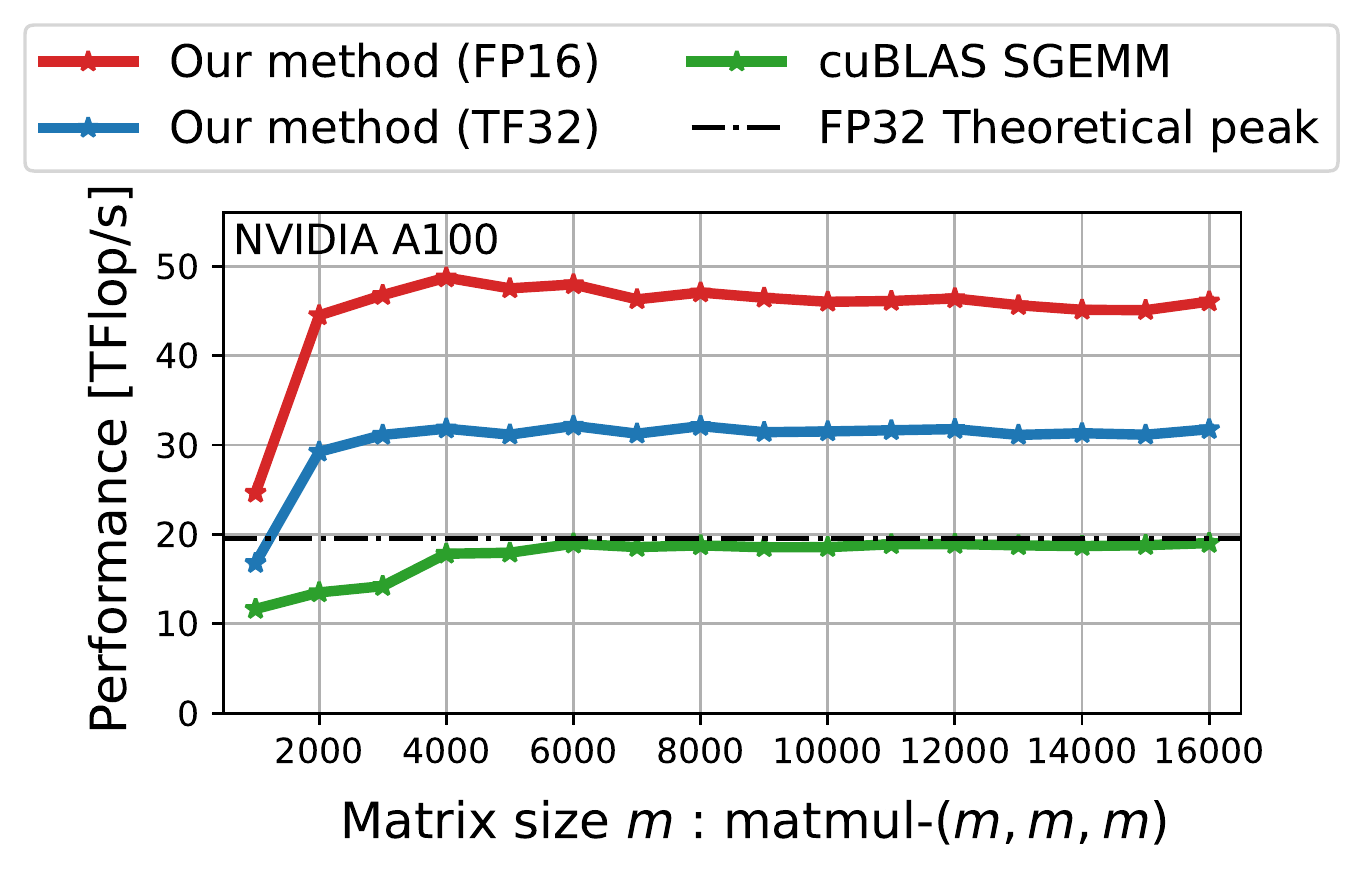}
    \caption{%
    Performance comparison of our method in TF32 and FP16, cuBLAS SGEMM and the FP32 theoretical peak.%
    Our methods compute the single-precision matrix-matrix multiplication with the same accuracy as cuBLAS SGEMM.
    }
    \label{fig:intro-performance}
\end{figure}

In this paper, we improve upon the existing error correction methods for matrix-multiplication on Tensor Cores by \cite{markidis_nvidia_2018} and \cite{feng_egemm-tc_2021}.
The two main causes of error in existing work are:
\begin{enumerate}
\item The rounding inside Tensor Cores is done with round-to-zero by default, which causes a large error even when accumulating in FP32.
\item The high probability of underflow and gradual underflow in the error correction computation.
\end{enumerate}
We address these problems and evaluate against four other existing methods; Markidis' method, Feng's method, cuBLAS SGEMM using FP32 SIMT Core, and cuBLAS SGEMM over Tensor Cores.
These results are shown in Figure \ref{fig:intro-residual}.
Although, we implemented the method described in Feng's work, we were not able to reproduce the accuracy shown in the paper \cite{feng_egemm-tc_2021}.
We also reduce the computational complexity compared to Markidis’s method and Feng's method.
Furthermore, we provide our SGEMM implementation using NVIDIA CUTLASS.\footnote{\url{https://github.com/enp1s0/cutlass}}
We evaluate the performance of our method in Figure \ref{fig:intro-performance}.
Our method surpasses the FP32 theoretical peak performance, while achieving the same error as FP32.

Our contributions can be summarized as follows:
\begin{itemize}
    \item
    We theoretically calculate the expectation of the mantissa length and experimentally evaluate the effect of this mantissa length.
    As a result, we find that the mantissa loss is not the main cause of error during matrix-matrix multiplication in Markidis' method.
    \item
    We evaluate the effect of rounding inside Tensor Cores, and find that this is one of the main causes of error.
    We reduce the effect of these rounding errors by accumulating outside of the Tensor Cores, and improve the matrix multiplication accuracy to exactly match that of CUDA FP32 SIMT Cores.
    \item
    We also apply scaling to reduce the underflow probability.
    As a result, our method can deal with a wider range of values compared to Feng's and Markidis' method.
    Furthermore, we use the TF32 \cite{nvidia_corporation_nvidia_nodate-1} data type, which is available on Ampere architectures, and confirm that it can represent nearly the entire FP32 exponent range.
    \item
    We remove one of the three error-correction terms, and reduce the amount of computation on Tensor Cores to 75\% without loss of any accuracy.
    \item
    We include these modifications to NVIDIA CUTLASS and evaluate the matrix-matrix multiplication accuracy, computational throughput, and power consumption. Our implementation shows higher performance and lower power consumption compared to cuBLAS SGEMM on FP32 SIMT Cores while retaining the same accuracy.
    \end{itemize}

\section{Background}

\subsection{Rounding}
The rounding of floating-point numbers is the key to understanding the present contribution, so we will first describe the different types of rounding.
Let us consider a floating-point value with $\ell$ mantissa bits, and the cases where it is rounded to $n$ bits.
\begin{equation}
\label{eq:rounding}
m = \overbrace{\underbrace{m_{\ell-1} m_{\ell-2} \cdots m_{\ell-n}}_{n \text{ \ bit}} m_{\ell-n-1} \cdots m_{0}}^{\ell \text{ \ bit}}
\end{equation}
The different rounding modes described below are defined by a combination of two basic operations; rounding-up and rounding-down (truncation). Rounding-up adds $1$ to $m_{\ell-1} m_{\ell-2} \cdots m_{\ell-n}$, while rounding-down does nothing to $m_{\ell-1} m_{\ell-2} \cdots m_{\ell-n}$.
In this paper, we use three types of rounding modes:
\begin{description}[style=nextline]
\item[Round to Nearest; ties to even (RN)]
\begin{enumerate}[label=\arabic*)]
    \item[] %
    \item Truncate when $m_{l-n-1}$ equals 0.
    \item When $m_{l-n-1}$ equals 1:
    \begin{enumerate}
        \item Round-up when at least one of $m_{l-n-2}, \cdots, m_0$ is $1$.
        \item Truncate or round up so that $m_{l-n}$ becomes $0$ when all of $m_{l-n-2}, \cdots, m_0$ are $0$  (Ties to even).
    \end{enumerate}
\end{enumerate}
\item[Round to Nearest; ties to Away from zero (RNA)]
\begin{enumerate}[label=\arabic*)]
    \item[] %
    \item Truncate when $m_{l-n-1}$ equals $0$.
    \item Round-up when $m_{l-n-1}$ equals $1$.
\end{enumerate}
\item[Round toward Zero (RZ)]
Always truncate.
\end{description}

\begin{figure}
    \centering
    \includegraphics[width=0.6\linewidth]{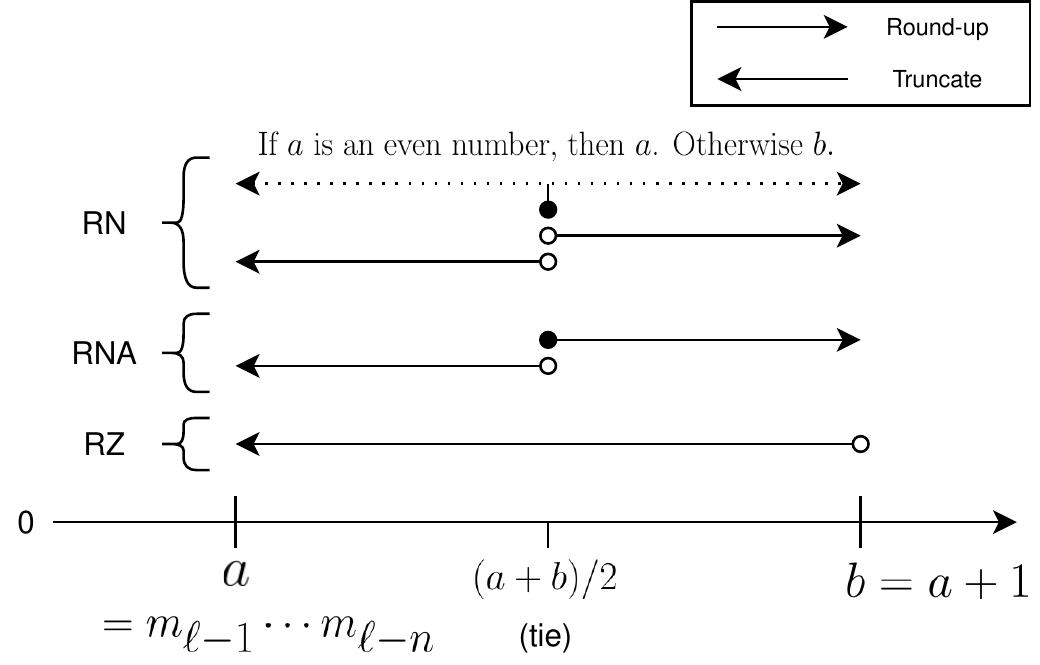}
    \caption{The roundings we use in this paper.}
    \label{fig:rounding}
\end{figure}

We show the number line representation of RN, RNA and RZ in Figure \ref{fig:rounding}.

\subsection{Tensor Cores}

\subsubsection{Programming interface}

NVIDIA provides the WMMA API for using Tensor Cores in CUDA/C++.
When using the WMMA API, the input matrices (in FP16) are copied to a register array called ``Fragments" using a WMMA API function.
A pseudocode for computing matrix-matrix multiplication $\mathbf{C} \leftarrow \mathbf{A}\times \mathbf{B}$ on Tensor Cores using WMMA API is shown in Code \ref{lst:stardard-wmma}.
In this pseudocode, we divide the input matrices into sub-matrices and compute matrix-matrix multiplications on them. The functions
 {\tt fill\_fragment}, {\tt load\_matrix\_sync}, {\tt mma\_sync}, {\tt mma\_sync} and {\tt store\_matrix\_sync} in the pseudocode are part of the WMMA API.
 The flow of computation is as follows:

\begin{enumerate}
    \item Initialize a fragment {frag\_c} for accumulating the resulting matrices using {\tt fill\_fragment} function.
    \item Load sub-matrices of $\mathbf{A}, \mathbf{B}$ from memory to fragments {\tt frag\_a}, {\tt frag\_b} using the {\tt load\_matrix\_sync} function.
    \item Compute the matrix-matrix multiplication of {\tt frag\_a}, {\tt frag\_b} and accumulate to {\tt frag\_c} using the {\tt mma\_sync} function.
    \item Store the sub-matrix of $\mathbf{C}$ from the fragment {\tt frag\_c} to memory.
\end{enumerate}

\begin{lstlisting}[style=CStyle,caption={A simple matrix-matrix multiplication pseudocode on Tensor Cores using WMMA API.},label={lst:stardard-wmma}]
__device__ void matmul(mem_c, mem_a[K], mem_b[K]) {
  fragment frag_a, frag_b, frag_c;
  shared_mem_fp16 smem_a, smem_b;
  // 1; Initialize an accumulator fragment
  fill_fragment(frag_c, 0.f);
  for (k=0;k<K;k++) {
    // Convert subdicided matrix to FP16
    smem_a = toFP16(mem_a[k]);
    smem_b = toFP16(mem_b[k]);
    // 2; Load subdivided matrices to fragments
    load_matrix_sync(frag_a, smem_a, ...);
    load_matrix_sync(frag_b, smem_b, ...);
    // 3; Compute matrix-matrix multiplication
    //    and accumulation on Tensor Cores
    mma_sync(frag_c, frag_a, frag_b, frag_c);
  }
  // 4; Store result to memory
  store_matrix_sync(mem_c, frag_c, ...);
}
\end{lstlisting}

\subsubsection{Fragment mapping and PTX Instructions}
The functions {\tt load\_matrix\_sync} and {\tt store\_matrix\_sync}, map the memory index of the input matrix elements to the fragment index.
We can analyze this mapping and use it for reducing the memory footprint.
Our {\tt wmma\_extension} library provides functions for that purpose. For instance, computing matrix-vector multiplication without unnecessary memory footprint, and loading/storing fragments with custom operations for each element.

There are two types of PTX instructions: {\tt wmma} and {\tt mma}.
The {\tt wmma} instructions provide finer control compared to the {\tt mma} instruction, such as loading fragments using {\tt wmma.load}, computing matrix-matrix multiplication accumulation using {\tt wmma.mma}, and storing fragments using {\tt wmma.store}.
However, we chose to use the {\tt mma} due to the more efficient register usage compared to {\tt wmma}.

When we use {\tt wmma} instructions, each element in matrices $\mathbf{A}$, and $\mathbf{B}$ is kept by two elements inside the fragments on threads in a warp.
On the other hand, for {\tt mma} each element of the input matrices is kept by only one element of the fragments on threads in a warp without duplication.
To use {\tt mma}, we need to map the memory and the fragment manually since the load and store instructions for {\tt mma} do not exist.
This map for {\tt mma} is available in the NVIDIA Toolkit Documentation\footnote{Warp Level Matrix Multiply-Accumulate Instructions section in PTX ISA Chapter \url{https://docs.nvidia.com/cuda/parallel-thread-execution/}}.

\subsection{Single-precision matrix-matrix multiplication using error correction technique on Tensor Cores}
As we mentioned previously, in order to use Tensor Cores for single-precision matrix-matrix multiplication we need to convert the input matrices to FP16 which introduces a truncation error.
Markidis \textit{et al.} propose a method to correct this truncation error by keeping the mantissa loss in additional FP16 variables and using them for correcting the accuracy of matrix-matrix multiplication \cite{markidis_nvidia_2018}.

\begin{eqnarray}
\label{eq:corr-1}
\mathbf{A}_\mathrm{F16} &\leftarrow& \mathrm{toFP16}\left( \mathbf{A}_\mathrm{F32} \right) \\
\label{eq:corr-2}
\Delta\mathbf{A}_\mathrm{F16} &\leftarrow& \mathrm{toFP16}\left( \mathbf{A}_\mathrm{F32} - \mathrm{toFP32}\left(\mathbf{A}_\mathrm{F16}\right)\right) \\
\label{eq:corr-3}
\mathbf{B}_\mathrm{F16} &\leftarrow& \mathrm{toFP16}\left( \mathbf{B}_\mathrm{F32} \right) \\
\label{eq:corr-4}
\Delta\mathbf{B}_\mathrm{F16} &\leftarrow& \mathrm{toFP16}\left( \mathbf{B}_\mathrm{F32} - \mathrm{toFP32}\left(\mathbf{B}_\mathrm{F16}\right)\right) \\
\hat{\mathbf{C}}_\mathrm{F32} &\leftarrow& \mathbf{A}_\mathrm{F16} \mathbf{B}_\mathrm{F16} + \Delta\mathbf{A}_\mathrm{F16} \mathbf{B}_\mathrm{F16}  \nonumber \\
\label{eq:corr-5}
&& + \mathbf{A}_\mathrm{F16} \Delta\mathbf{B}_\mathrm{F16} + \Delta\mathbf{A}_\mathrm{F16} \Delta\mathbf{B}_\mathrm{F16}
\end{eqnarray}

\begin{lstlisting}[style=CStyle,caption={A simple pseudocode example of Markidis' error correction method},label={lst:markidis-ec}]
void matmul(mem_c, mem_a[K], mem_b[K]) {
  fragment frag_a, frag_b, frag_c;
  fragment frag_da, frag_db;
  shared_mem_fp16 smem_a, smem_b;
  shared_mem_fp16 smem_da, smem_db;
  // Initialize an accumulator fragment
  fill_fragment(frag_c,0.f);
  for (k=0;k<K;k++) {
    // Compute eq 2, 4
    smem_a=toFP16(mem_a[k])
    smem_b=toFP16(mem_b[k])
    //
    load_matrix_sync(frag_a,smem_a);
    load_matrix_sync(frag_b,smem_b);
    // Compute eq 3, 5
    smem_da=toFP16(mem_a[k]-toFP32(smem_a))
    smem_db=toFP16(mem_b[k]-toFP32(smem_b))
    //
    load_matrix_sync(frag_da,smem_da);
    load_matrix_sync(frag_db,smem_db);
    // Compute eq 6 and accumulate
    mma_sync(frag_c,frag_da,frag_db,frag_c);
    mma_sync(frag_c,frag_da,frag_b,frag_c);
    mma_sync(frag_c,frag_a,frag_db,frag_c);
    mma_sync(frag_c,frag_a,frag_b,frag_c);
  }
  // Store result to memory
  store_matrix_sync(mem_c,frag_c);
}
\end{lstlisting}

We show the accuracy of Markidis' method in Figure \ref{fig:intro-residual}.
To evaluate the accuracy, we compute the relative residual using the following equation.
\begin{equation}
\label{eq:eval-matmul}
\mathrm{RelativeResidual} = \frac{||\mathbf{C}_\mathrm{FP64} - \mathbf{C}_\mathrm{Target}||_F}{||\mathbf{C}_\mathrm{FP64}||_F},
\end{equation}
where the $||\cdot||_F$ is the Frobenius norm.
We compute the reference matrix-matrix multiplication $\mathbf{C}_\mathrm{FP64} = \mathrm{toFP64}\left(\mathbf{A}_\mathrm{FP32}\right) \cdot\mathrm{toFP64}\left(\mathbf{B}_\mathrm{FP32}\right)$ in FP64.

From Figure \ref{fig:intro-residual} we see that, Markidis' method is more accurate than the Tensor Core without error correction for smaller matrix sizes.
However, we found that RZ rounding inside Tensor Cores accumulates errors faster than the RN of FP32 SIMT Cores, and the accuracy catches up to the Tensor Core as the matrix size becomes larger.
Feng \textit{et al.} propose to reduce the mantissa loss in Eqs. (\ref{eq:corr-2}) and (\ref{eq:corr-4}) by using the sign bit as an extra mantissa bit \cite{feng_egemm-tc_2021}.
However, our experiments could not show any advantage by faithfully reproducing what is described in their paper.
\section{Error investigation and improvements}
\subsection{Expectation of mantissa length}
\label{sec:mantissa-length-discussion}
We can write Eqs. (\ref{eq:corr-1}) and (\ref{eq:corr-2}) and Eqs. (\ref{eq:corr-3}) and (\ref{eq:corr-4}) for each element as
\begin{eqnarray}
\label{eq:corr-elm-1}
v_\mathrm{F16} &\leftarrow& \mathrm{toFP16}\left( v_\mathrm{F32} \right) \\
\label{eq:corr-elm-2}
\Delta v_\mathrm{F16} &\leftarrow& \mathrm{toFP16}\left( v_\mathrm{F32} - \mathrm{toFP32}\left(v_\mathrm{F16}\right)\right).
\end{eqnarray}
The input element $v_\text{F32}$ is approximated by $v_\text{F16} + \Delta v_\text{F16}$.
The mantissa length of FP32 is $23 + 1 = 24$ bit, including an implicit $1$ bit, and the FP16 is $10 + 1 = 11$ bit.
Thus $v_\mathrm{F16} + \Delta v_\mathrm{F16}$ can not represent the full mantissa of $v_\mathrm{F32}$.
In this section, we calculate the expectation of the mantissa length kept by Eq. (\ref{eq:corr-elm-1}) and (\ref{eq:corr-elm-2}).

We express the mantissa bit of $v_\mathrm{F32}$ as $m_{22} m_{21} \cdots m_{0}$ from MSB without the implicit 1 bit,
and the rounding during the conversion to FP16 is RN, which is the default in CUDA.
In this case, $m_{13}\cdots m_{0}$ bits decide whether we round-up in Eq. (\ref{eq:corr-elm-1}).
We show the mantissa length kept by Eqs. (\ref{eq:corr-elm-1}) and (\ref{eq:corr-elm-2}) and its probability of occurrence in the case of $m_{13}\cdots m_{0}$ in Table \ref{tab:halfhalf-mantissa}.
This probability is calculated under Assumption \ref{ass:iid} for the mantissa part of floating-point numbers as follows.

\begin{assumption}
\label{ass:iid}
Each bit of the mantissa is 1 with probability $\frac{1}{2}$, and each bit is statistically independent
\end{assumption}

It follows that the expectation of the mantissa length is 22.75 bits out of the FP32 mantissa length of 23 bits.
Furthermore, when we use RNA for rounding during the FP16 conversion in Eqs. (\ref{eq:corr-elm-1}) and (\ref{eq:corr-elm-2}), the values of $v_\mathrm{F16}$ and $\Delta v_\mathrm{F16}$ are different from RN.
However, the mantissa length and its probability of occurrence are the same as RN, and the expectation of mantissa length is 22.75 bits.

\begin{figure}[t]
    \centering
    \includegraphics[width=0.6\linewidth]{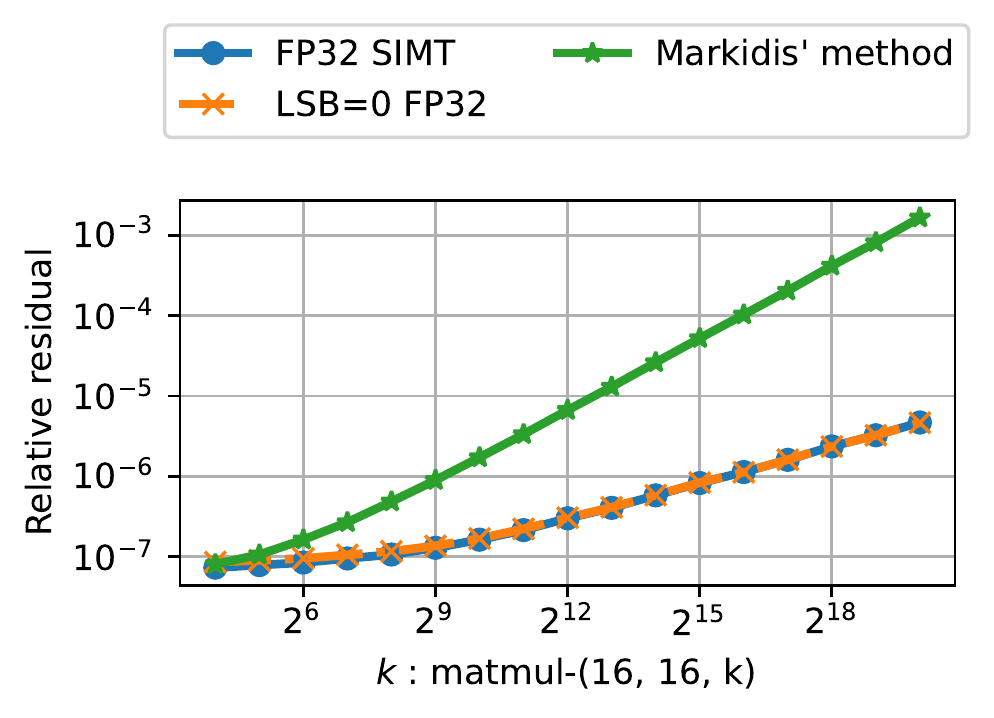}
    \caption{The accuracy comparison between Markidis' method, FP32 SIMT Core, and truncating the last $1$ bit of FP32 mantissa. Input matrices $\mathbf{A}$ and $\mathbf{B}$ are initialized with a uniform distribution $(-1, 1)$.}
    \label{fig:mantissa22}
\end{figure}

To evaluate the effect of this 22.75 bits of mantissa, we evaluate the accuracy of FP32 matrix-matrix multiplication by truncating the Least Significant Bit (LSB) of the mantissa of input matrices.
The expectation of the mantissa length during this operation is 22.5 bits under Assumption \ref{ass:iid}, which is shorter than 22.75 bits.
We show the accuracy comparison between this truncation and Markidis' method in Figure \ref{fig:mantissa22}, and see that the accuracy of Markidis' method is worse than this method despite the higher expectation of mantissa length.
In conclusion, the mantissa loss is not the main cause of error during matrix-matrix multiplication on Tensor Cores.

\begin{table}[]
\center
\begin{tabular}{llllll|ll}
\toprule
$l_0$                 & $m_{13}$ & $m_{12}$ & $m_{11}$ & $m_{1}$ & $m_{0}$ & len & prob \\
\midrule
$\geq 2$              & *        & 0        & 0        & *       & *       & 23     & 1/4         \\
\midrule
\multirow{2}{*}{$=1$} & *        & 0        & 1        & *       & 0       & 23     & 1/8         \\
                      & *        & 0        & 1        & *       & 1       & 22     & 1/8         \\
\midrule

\multirow{3}{*}{$=0$} & *        & 1        & 0        & *       & 1       & 22     & 1/8        \\
                      & *        & 1        & 0        & *       & 0       & 23     & 1/8        \\
                      & *        & 1        & 1        & *       & *       & 23     & 1/4   \\
\bottomrule
\end{tabular}
\caption{The mantissa length kept by $v_\mathrm{F16}$ and $\Delta v_\mathrm{F16}$ (len), the probability of occurrence (prob) when RN is performed during the FP16 conversion in Eqs. (\ref{eq:corr-elm-1}) and ( \ref{eq:corr-elm-2}). $m_{22} m_{21} \cdots m_{0}$ represents the 23 bits of FP32 mantissa, and the probability is calculated under Assumption \ref{ass:iid}. The length $l_0$ is the number of consecutive zeros from $m_{12}$ towards the LSB. The mark "*" means it does not matter if it is 0 or 1.}
\label{tab:halfhalf-mantissa}
\end{table}

Although the mantissa length of FP16 is $10 + 1 = 11$ bits and two FP16s can keep only $22$ bits of mantissa per $23 + 1 = 24$ bits of FP32 mantissa, the expectation of the mantissa length is $22.75 + 1 = 23.75$ bits.
The reason for this can be explained as follows:
\begin{itemize}
    \item
    When the last $n$ bits of mantissa are $0$, the mantissa length to keep is $23 + 1 - n$ bits per 24 bits.
    We can keep the full mantissa when $n \geq 2$.
    \item
    When $l_0$ is greater than or equal to $2$, the mantissa length kept by $\Delta v_\text{F16}$ is less than or equal to $10$.
    Therefore, we can keep the full mantissa.
    \item
Rounding-up can be performed during the conversion in Eq. (\ref{eq:corr-elm-1}) when $l_0 = 0$.
It keeps more mantissa compared to RZ.
It follows that $|v_\mathrm{FP16}| > |v_\mathrm{FP32}|$ when rounding-up.
And the signs of $\Delta v_\mathrm{FP16}$ and $v_\mathrm{FP32}$ are different because $v_\mathrm{FP16}$ and $v_\mathrm{FP32}$ have the same sign.
Let us consider the following example.
The mantissa of $v_\mathrm{FP32}$ is represented by a $23+1=24$ bit integer shown in Eq. (\ref{eq:mantissa-example}).
\begin{equation}
    \label{eq:mantissa-example}
    I_\mathbf{FP32} = +\underbrace{10000000000'}_{\mathbf{11bit}}\underbrace{10000000000'}_{\mathbf{11bit}}11
\end{equation}
The $I_\text{FP32}$ is kept by two $10 + 1 = 11$ bit integers $I_\mathrm{FP16}$ and $\Delta I_\mathrm{FP16}$ using RZ.
\begin{alignat*}{3}
    &I_\mathbf{FP16}        &=& +\underbrace{10000000000'}_{\mathbf{11bit}} &                                             &                                   \\
    &\Delta I_\mathbf{FP16} &=&                                             & +\underbrace{10000000000'}_{\mathbf{11bit}} &                                   \\
    &\mathrm{Loss}          &=&                                             &                                             & +\underbrace{11}_{\mathbf{2bit}}
\end{alignat*}
In this case, 2 bits of mantissa loss occurs.
On the other hand, when we use RN instead of RZ, only 1 bit of mantissa loss occurs.
\begin{alignat*}{3}
    &I_\mathbf{FP16}        &=& +\underbrace{10000000001'}_{\mathbf{11bit}} &                                               &                                       \\
    &\Delta I_\mathbf{FP16} &=&                                             & -0\underbrace{1111111111'0}_{\mathbf{11bit}}  &                                       \\
    &\mathrm{Loss}          &=&                                             &                                               & +\underbrace{1}_{\mathbf{1bit}}
\end{alignat*}
This is because we need fewer bits to keep $n$ bits of an integer $a$ when $a > 2^{n-1}$, if we keep $2^n - a$ instead of $a$.
\end{itemize}

\begin{table}[]
\center
\begin{tabular}{llllll|ll}
\toprule
$l_0$                 & $m_{13}$ & $m_{12}$ & $m_{11}$ & $m_{1}$ & $m_{0}$ & len & prob \\
\midrule
$\geq 3$              & *        & 0        & 0        & *       & *       & 23     & 1/4         \\
\midrule
\multirow{2}{*}{$=2$} & *        & 0        & 1        & *       & 0       & 23     & 1/8         \\
                      & *        & 0        & 1        & *       & 1       & 22     & 1/8         \\
\midrule

\multirow{3}{*}{$=1$} & *        & 1        & *        & 1       & *       & 21     & 1/4        \\
                      & *        & 1        & *        & 0       & 1       & 22     & 1/8        \\
                      & *        & 1        & *        & 0       & 0       & 23     & 1/8        \\
\bottomrule
\end{tabular}
\caption{The mantissa length kept by $v_\mathrm{F16}$ and $\Delta v_\mathrm{F16}$ (len), the probability of occurrence (prob) when RZ is performed during the FP16 conversion in Eqs. (\ref{eq:corr-elm-1}) and  (\ref{eq:corr-elm-2}). $m_{22} m_{21} \cdots m_{0}$ represents the 23 bits of FP32 mantissa, and the probability is calculated under Assumption \ref{ass:iid}. $l_0$ is the number of consecutive zeros from $m_{12}$ toward LSB. The mark "*" means it does not matter if it is 0 or 1.}
\label{tab:halfhalf-mantissa-rz}
\end{table}

We also calculate the expectation of the mantissa length when we use RZ in Eqs. (\ref{eq:corr-elm-1}) and (\ref{eq:corr-elm-2}).
We show the mantissa length and its probability of occurrence in Table \ref{tab:halfhalf-mantissa-rz}.
The expectation of the mantissa length is 22.25 bits.

\subsection{Avoiding RZ during Tensor Core accumulation}
The accumulator inside Tensor Cores has at least 2 extra bits of mantissa and RZ is used for rounding \cite{fasi_numerical_2020}.
It follows that RZ is performed in the accumulator {\tt frag\_c} in every $k$ iteration in Code \ref{lst:markidis-ec}.
We evaluate the effect of this RZ for the matrix-matrix multiplication using Markidis' method using the mixed-precision matrix-matrix multiplication function {\tt mma\_rn} and {\tt mma\_rz} that perform similar operations with Tensor Cores.
Both functions compute the matrix-matrix multiplication as follows
\begin{equation}
    \label{eq:matmul-0}
    \mathbf{D}_\mathrm{F32} \leftarrow \mathbf{A}_\mathrm{F16} \times \mathbf{B}_\mathrm{F16} + \mathbf{C}_\mathrm{F32}.
\end{equation}
The multiplication of each element is computed in FP32 and accumulation in FP64.
We truncate the mantissa of the accumulator to keep them in 25 bit after every element accumulation.
The difference between these two functions is that the {\tt mma\_rn} performs RN for rounding after the addition of $\mathbf{C}_\mathrm{F32}$ in Eq. (\ref{eq:matmul-0}), while the {\tt mma\_rz} performs RZ in the same way as Tensor Cores do.

\begin{figure}[t]
    \centering
    \includegraphics[width=0.6\linewidth]{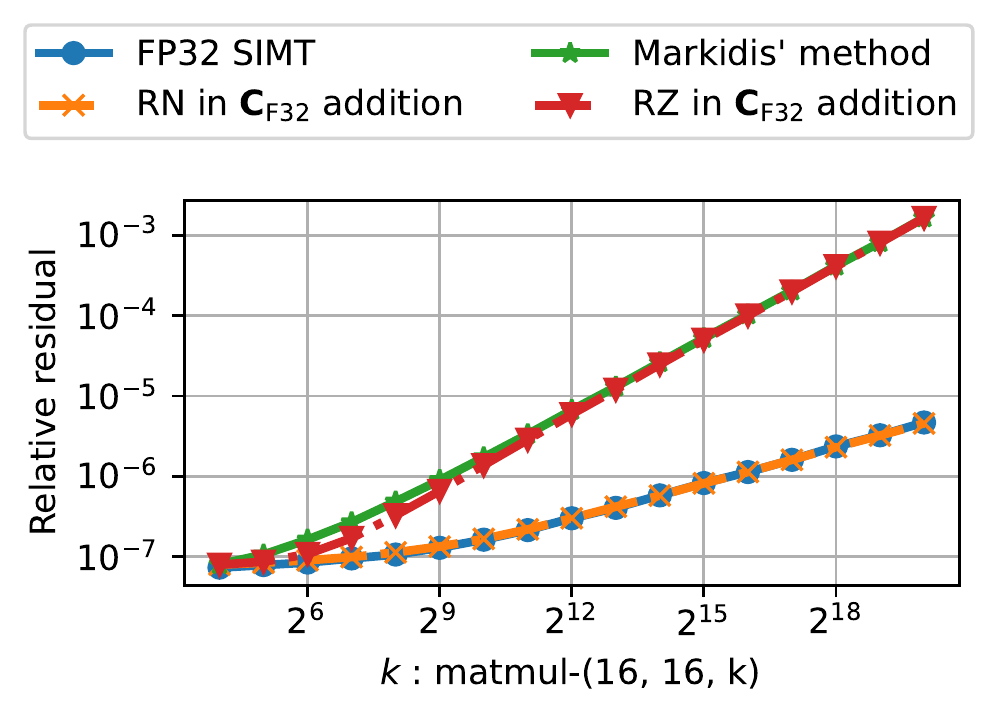}
    \caption{The evaluation of the accuracy of single-precision matrix-matrix multiplication using Markidis' error correction method with {\tt mma\_rn} and {\tt mma\_rz} instead of Tensor Cores. %
    }
    \label{fig:tc-emu-accuracy}
\end{figure}

We use {\tt mma\_rz} and {\tt mma\_rn}, instead of Tensor Cores, to compute a single-precision matrix-matrix multiplication using Markidis' method and evaluate the accuracy.
The results are shown in Figure \ref{fig:tc-emu-accuracy}.
While the accuracy using {\tt mma\_rn} is the same as FP32 SIMT Core, the one using {\tt mma\_rz} is the same as Markidis' method.
Therefore, we conclude that performing RZ in the accumulator after addition to $\mathbf{C}_\text{F32}$ in Eq. (\ref{eq:matmul-0}) causes the accuracy loss in Markidis' method.

\begin{figure}[t]
    \centering
    \includegraphics[width=0.6\linewidth]{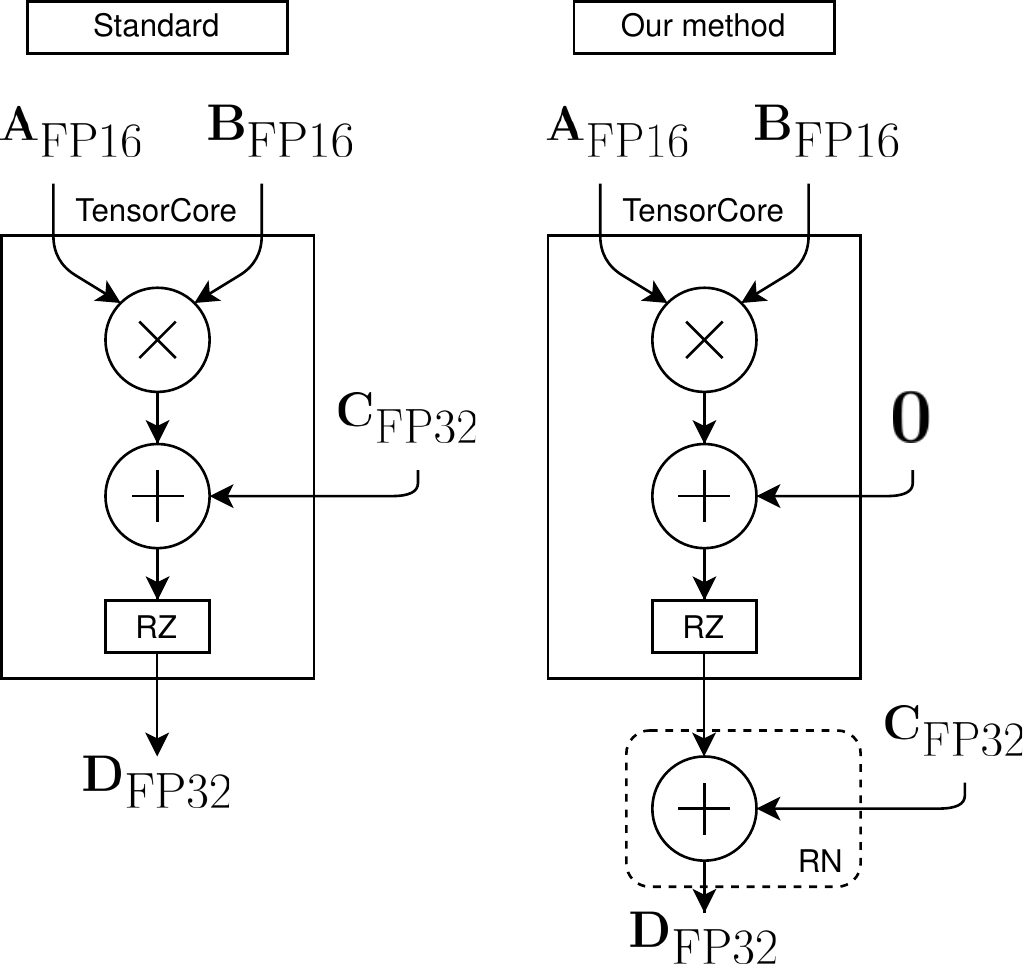}
    \caption{The flow of computation to avoid RZ inside Tensor Cores, which affects the accuracy of Markidis' error correction method. %
    {\bf Left}: Standard usage of Tensor Cores via WMMA API. %
    In this case, RZ is performed on the accumulator $\mathbf{C}_\mathrm{FP32}$ directly. %
    {\bf Right}: Our method for avoiding the RZ. We accumulate the result of the matrix-matrix multiplication $\mathbf{A}_\text{F16} \times \mathbf{B}_\text{F16}$ to $\mathbf{C}_\text{F32}$ outside of Tensor Cores using FP32 SIMT Cores.
    }
    \label{fig:avoid-tc-rounding}
\end{figure}

To avoid the RZ and improve the accuracy, we compute the addition to $\mathbf{C}_\text{F32}$ in Eq. (\ref{eq:matmul-0}) outside of the Tensor Cores.
We show the flow of our proposed method, and compare it against the standard process presented in Figure \ref{fig:avoid-tc-rounding}.
We input a zero matrix to the Tensor Cores and compute the addition shown in Eq. (\ref{eq:matmul-0}) outside of the Tensor Cores using FP32 SIMT Cores which performs RN for rounding.
By using this technique, the accuracy of the single-precision matrix-matrix multiplication using Markidis' method improves to the same accuracy as FP32 SIMT Cores, as shown in Figure \ref{fig:intro-residual}.
On the other hand, this technique requires more registers to keep a zero matrix, additional process for making the zero matrix, and extra addition operations on FP32 SIMT Cores compared to Markidis' method.

\subsection{Reducing the underflow and gradual underflow probability in $\Delta\mathbf{v}_\mathrm{F16}$ computations}

\begin{figure}[t]
    \centering
    \includegraphics[width=0.6\linewidth]{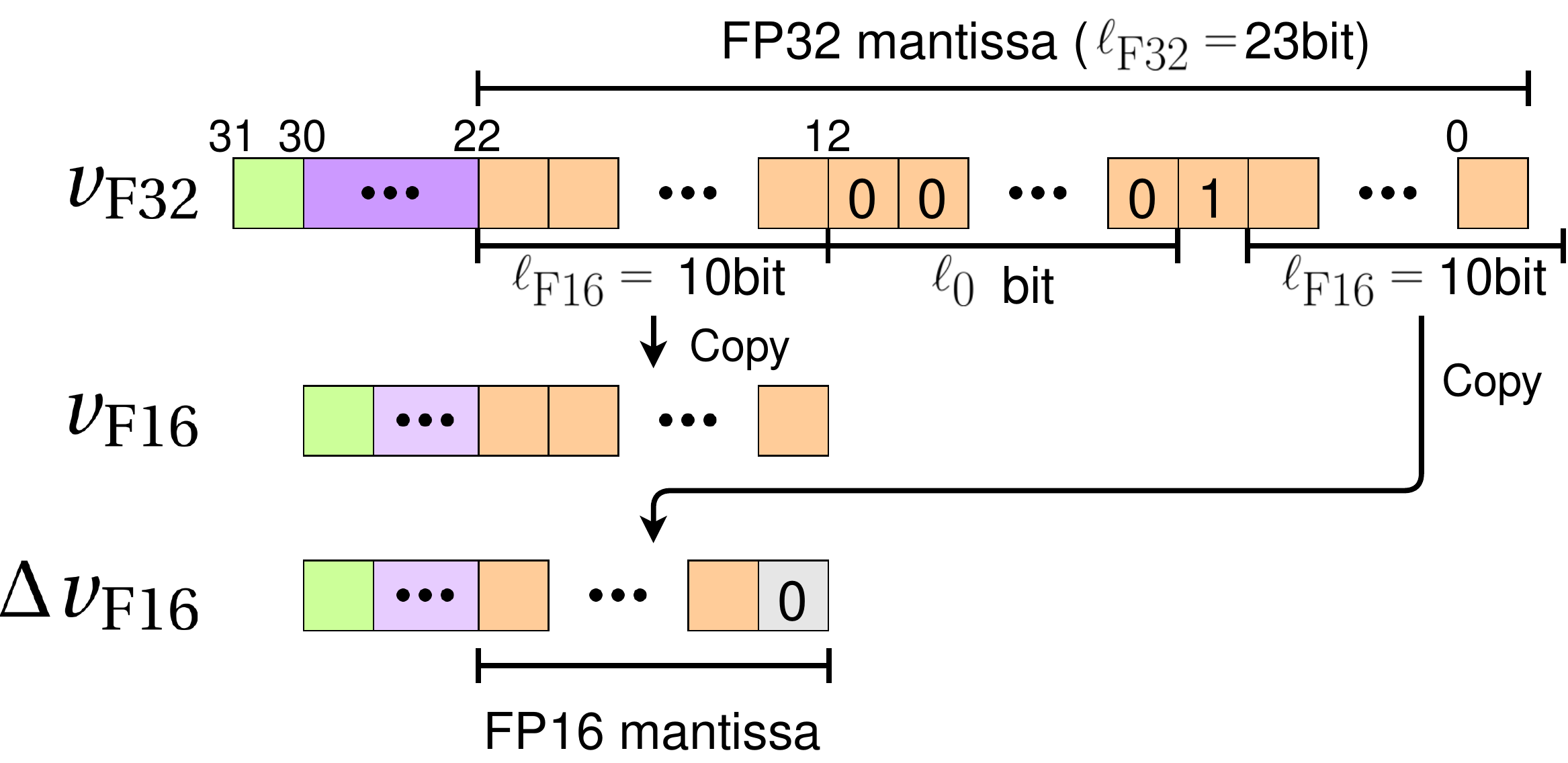}
    \caption{An example of the mantissa bitstring movement in Eqs. (\ref{eq:corr-elm-1}) and (\ref{eq:corr-elm-2}) with RZ. When $l_0$ is larger than 2, the last $l_0 - 2$ bits of the mantissa of $\Delta v_\text{F16}$ are filled with zeros.}
    \label{fig:halfhalf-underflow}
\end{figure}

The probabilities of underflow and gradual underflow in subtracting two values are high when their absolute values are close.
This is shown in Eq. (\ref{eq:corr-elm-2}).
We calculate the probabilities for each exponent value of $v_\text{F32}$ and improve Eq. (\ref{eq:corr-elm-2}) to reduce the underflow.
To calculate these probabilities, we first define some constant values, the exponent bias of FP16 $b_\text{F16} = 15$, and the mantissa length of FP16 $l_\text{F16} = 10$ and FP32 $l_\text{F32} = 23$.
To simplify the calculation, we assume that RZ is used in toFP16 in Eqs. (\ref{eq:corr-elm-1}) and (\ref{eq:corr-elm-2}) while RN is used otherwise.
Under this assumption, the first 10 bits of the mantissa of $v_\mathrm{F32}$ are kept by $v_\mathrm{F16}$, and the 10 bits from $(l_\text{F16} + l_0)$th bit are kept by $\Delta v_\mathrm{F16}$ shown in Figure \ref{fig:halfhalf-underflow}.
Therefore, the exponent value of $\Delta v_\mathrm{F16}$ is smaller than the $v_\mathrm{F32}$ by $l_0 + l_\text{F16} + 1$, the last 1 is the length of the implicit bit of the mantissa.
We calculate the probabilities using this fact into consideration.
We also define the exponent value of $v_\mathrm{F32}$ as $e_v$ including exponent bias, where $v_\mathrm{F32}$ is represented as follows,
\begin{equation}
\label{eq:v32-represent}
v_\mathrm{F32}= 1.m_{22}m_{21}\cdots m_{0} \times 2^{e_v}
\end{equation}

First, we calculate $P_{u+gu}(e_v)$, the probability at which gradual underflow occurs.
When normalized, the smallest number that can be expressed in FP16 is $2^{-b_\text{F16} + 1}$.
Therefore, the condition for underflow or gradual underflow in Eq. (\ref{eq:corr-elm-2}) can be represented as
\begin{eqnarray}
    e_v - (l_0 + l_\text{F16} + 1) &<& - b_\text{F16} + 1 \nonumber\\
    \label{eq:v_gunderflow_cond}
    \Rightarrow e_v - l_\text{F16} + b_\text{F16} -2 &<& l_0.
\end{eqnarray}
We define the probability $P(l_0=n)$ such that $l_0$ equals to $n$ under the assumption \ref{ass:iid} as
\begin{equation}
    \label{eq:up}
    P(l_0 = n) =
    \begin{cases}
        0 & (n < 0) \\
        {\displaystyle\left(\frac{1}{2}\right)^{n + 1}} & (0 \leq n < l_\text{F32} - l_\text{F16}) \\
        {\displaystyle\left(\frac{1}{2}\right)^{l_\text{F32} - l_\text{F16}}} & (n = l_\text{F32} - l_\text{F16})
    \end{cases}
\end{equation}
Then, by using $P(l_0 = n)$, we calculate the desired probability $P_{u+gu}(e_v)$ as
\begin{equation}
    \label{eq:prob_g_u}
    P_{u+gu}(e_v) =
        {\displaystyle \sum_{l=(e_v-l_\text{F16}+b_\text{F16} - 2) + 1}^{l_\text{F32} - l_\text{F16}}} P(l_0 = l).
\end{equation}

Next, we calculate $P_{u}(e_v)$, the probability where only underflow occurs.
Without normalization, the smallest number that can be expressed by FP16 is $2^{-(b_\text{F16} + l_\text{F16}) + 1}$.
Therefore, the condition at which underflow occurs can be expressed as
\begin{eqnarray}
    e_v - (l_0 + l_\text{F16} + 1) &<& - (b_\text{F16} + l_\text{F16}) + 1  \nonumber\\
    \Rightarrow e_v + b_\text{F16} - 2 &<& l_0
    \label{eq:v_underflow_cond}
\end{eqnarray}
Using $P(l_0 = n)$ in the same way as $P_{u+gu}(e_v)$, we can calculate the desired probability $P_{u}(e_v)$ as
\begin{equation}
    \label{eq:prob_u}
    P_{u}(e_v) =
        {\displaystyle \sum_{l=(e_v + b_\text{F16} - 2) + 1}^{l_\text{F32} - l_\text{F16}}} P(l_0 = l)
\end{equation}

\begin{figure}[t]
    \centering
    \includegraphics[width=0.6\linewidth]{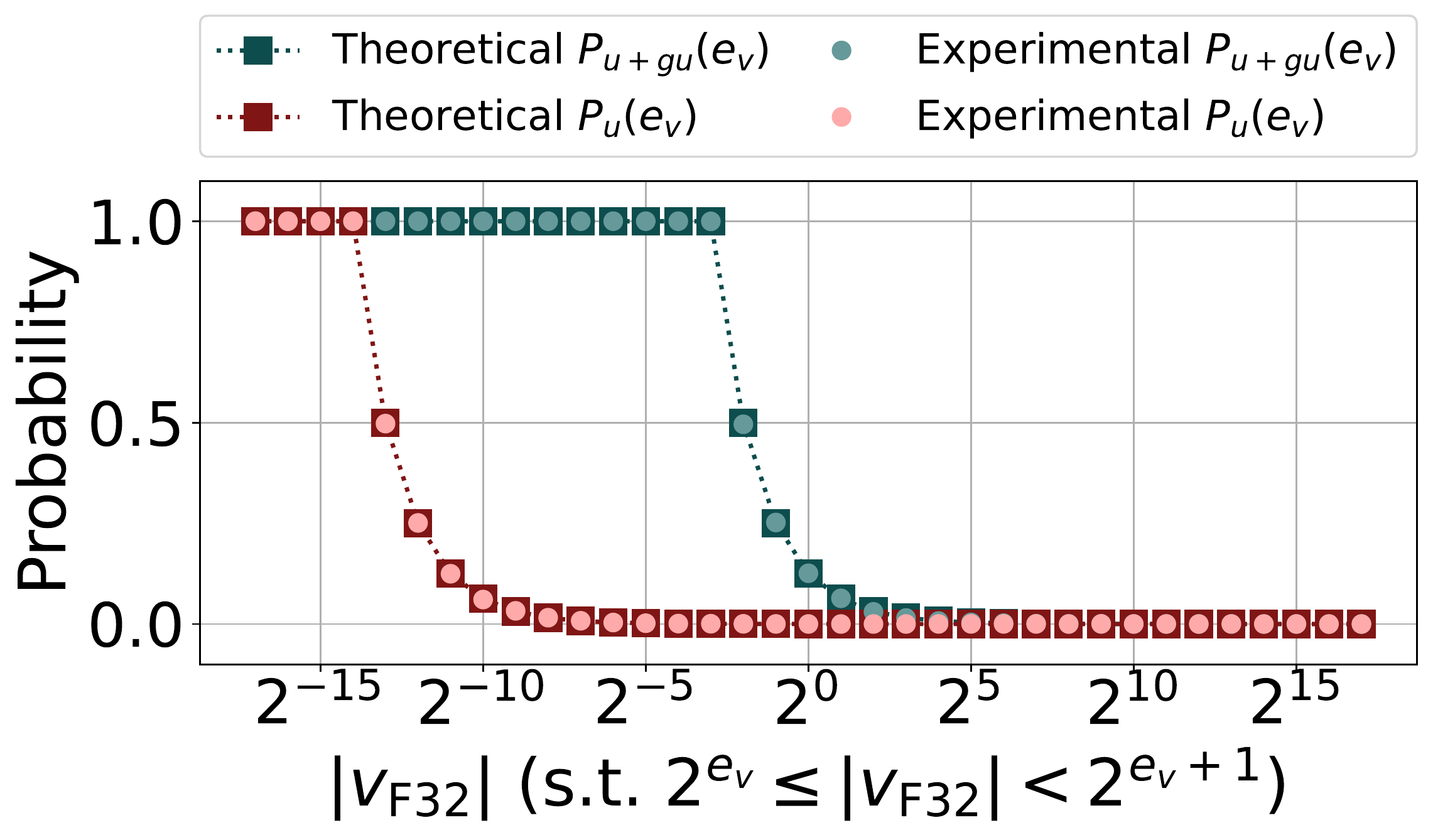}
    \caption{The theoretical and experimental probability of underflow $P_\text{u}(e_v)$ in Eq. (\ref{eq:corr-elm-2}) and the sum of underflow and gradual underflow $P_\text{u+gu}(e_v)$.}
    \label{fig:gu_prob}
\end{figure}

In Figure \ref{fig:gu_prob}, we show the theoretical $P_{u + gu}(e_v)$ and $P_{u}(e_v)$ calculated by Eq. (\ref{eq:prob_g_u}) and Eq. (\ref{eq:prob_u}) respectively, along with the experimental values aggregated on GPUs.
We find that gradual underflow occurs in Eq. (\ref{eq:corr-elm-2}) even if $v_\text{F32}$ is around $10^0$.

To reduce these probabilities, we add $l_\text{F16} + 1 = 11$ to the exponent of the result of subtraction in Eq. (\ref{eq:corr-elm-2}) by multiplying $2^{11}$,
\begin{eqnarray}
\Delta v_\mathrm{F16} &\leftarrow& \mathrm{toFP16}\left(\left( v_\mathrm{F32} - \mathrm{toFP32}\left(v_\mathrm{F16}\right)\right) \times 2^{11}\right) \nonumber \\
\label{eq:corr-elm-2-1024}
\end{eqnarray}
This process does not affect the mantissa.
In this paper, we refer to the method that keeps $v_\mathrm{F32}$ in Eqs. (\ref{eq:corr-elm-1}) and (\ref{eq:corr-elm-2-1024}) as {\bf halfhalf}, and the method in Eqs. (\ref{eq:corr-elm-1}) and (\ref{eq:corr-elm-2}) as {\bf Markidis' halfhalf}.
The single-precision matrix-matrix multiplication using halfhalf can be written as
\begin{eqnarray}
\label{eq:corr-1-1024}
\mathbf{A}_\mathrm{F16} &\leftarrow& \mathrm{toFP16}\left( \mathbf{A}_\mathrm{F32} \right) \\
\Delta\mathbf{A}_\mathrm{F16} &\leftarrow& \mathrm{toFP16}\left( \left(\mathbf{A}_\mathrm{F32} - \mathrm{toFP32}\left(\mathbf{A}_\mathrm{F16}\right)\right) \times 2^{11}\right) \nonumber \\
\label{eq:corr-2-1024}
\\
\label{eq:corr-3-1024}
\mathbf{B}_\mathrm{F16} &\leftarrow& \mathrm{toFP16}\left( \mathbf{B}_\mathrm{F32} \right) \\
\Delta\mathbf{B}_\mathrm{F16} &\leftarrow& \mathrm{toFP16}\left( \left( \mathbf{B}_\mathrm{F32} - \mathrm{toFP32}\left(\mathbf{B}_\mathrm{F16}\right)\right) \times 2^{11} \right) \nonumber \\
\label{eq:corr-4-1024}
\\
\hat{\mathbf{C}}_\mathrm{F32} &\leftarrow& \mathbf{A}_\mathrm{F16} \mathbf{B}_\mathrm{F16} \nonumber \\
&& + \left(\Delta\mathbf{A}_\mathrm{F16} \mathbf{B}_\mathrm{F16}  + \mathbf{A}_\mathrm{F16} \Delta\mathbf{B}_\mathrm{F16}\right) /2^{11} \nonumber \\
\label{eq:corr-5-1024}
&& + \left(\Delta\mathbf{A}_\mathrm{F16} \Delta\mathbf{B}_\mathrm{F16}\right)/ 2^{22}
\end{eqnarray}

\begin{figure}[t]
    \centering
    \includegraphics[width=0.6\linewidth]{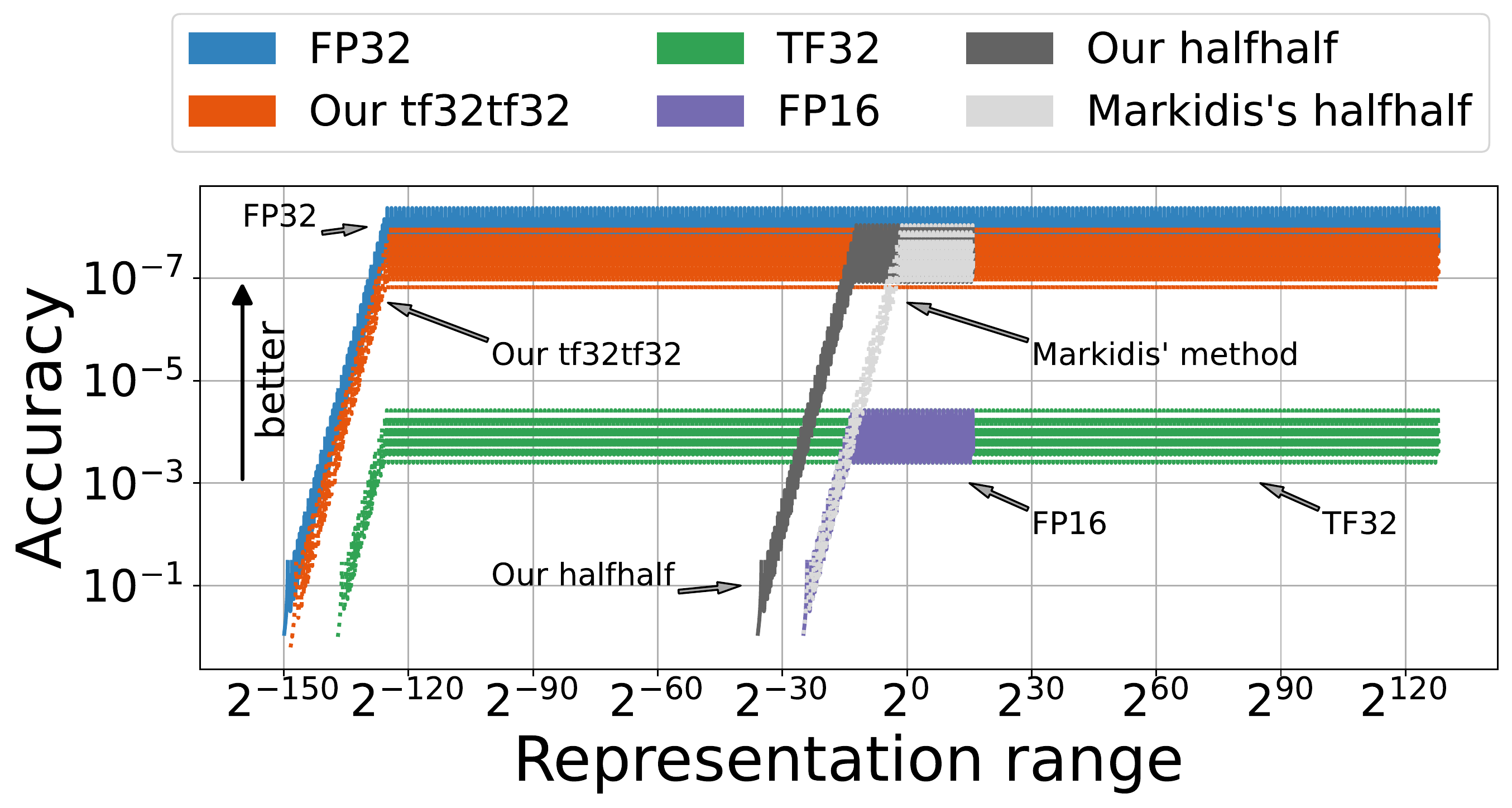}
    \caption{The comparison of representation accuracy and exponent range among FP32, FP16, TF32, halfhalf, tf32tf32, and Markidis' halfhalf.}
    \label{fig:halfhalf-precision}
\end{figure}

CUDA provides a data type called TF32 (Tensor Float) which has $8$ bits of exponent and $10$ bits of mantissa as the input type of Tensor Cores in Ampere architectures.
Because the exponent length is the same as FP32, we can keep a wider exponent range compared to FP16.
We use the TF32 instead of FP16 in Eqs. (\ref{eq:corr-elm-1}) and (\ref{eq:corr-elm-2}) and we refer to this method as {\bf tf32tf32}.
Currently, we can use RNA and RZ for rounding when converting FP32 to TF32.
We use RNA because it keeps more mantissa compared to RZ, as we have shown in the Section {\it Expectation of mantissa length}.
We show the comparison of representation accuracy and exponent range in Figure \ref{fig:halfhalf-precision}.

\subsection{Removing negligible error correction}
The absolute value of each element in $\Delta^{(2)} = \Delta\mathbf{A}_\mathrm{F16} \Delta\mathbf{B}_\mathrm{F16}$ is at least $2^{22}$ times smaller than $\Delta^{(0)} = \mathbf{A}_\mathrm{F16} \mathbf{B}_\mathrm{F16}$.
When two floating-point values are added, the mantissa of the value with the smaller exponent is shifted to align the exponent of the two values.
Therefore, when computing $\Delta^{(0)} + \Delta^{(2)}$, the shifting size is at least $22$ bits in each element.
This means that each $\Delta^{(2)}$ element at most affects the mantissa LSB in each $\Delta^{(0)}$ element since the mantissa length of FP32 is $23$ bit and the correction capability is negligible.
Thus, we ignore this term and replace Eq. (\ref{eq:corr-5-1024}) with Eq. (\ref{eq:corr-5-1024-reduce}).
\begin{eqnarray}
\hat{\mathbf{C}}_\mathrm{F32} &\leftarrow& \mathbf{A}_\mathrm{F16} \mathbf{B}_\mathrm{F16} \nonumber \\
&& + \left(\Delta\mathbf{A}_\mathrm{F16} \mathbf{B}_\mathrm{F16}  + \mathbf{A}_\mathrm{F16} \Delta\mathbf{B}_\mathrm{F16}\right) /2^{11} \nonumber \\
\label{eq:corr-5-1024-reduce}
\end{eqnarray}
Eq. (\ref{eq:corr-5-1024-reduce}) reduces the computation on Tensor Cores by 3/4.

\subsection{Incorporating our method into CUTLASS}
\begin{figure*}
    \centering
    \includegraphics[width=\linewidth]{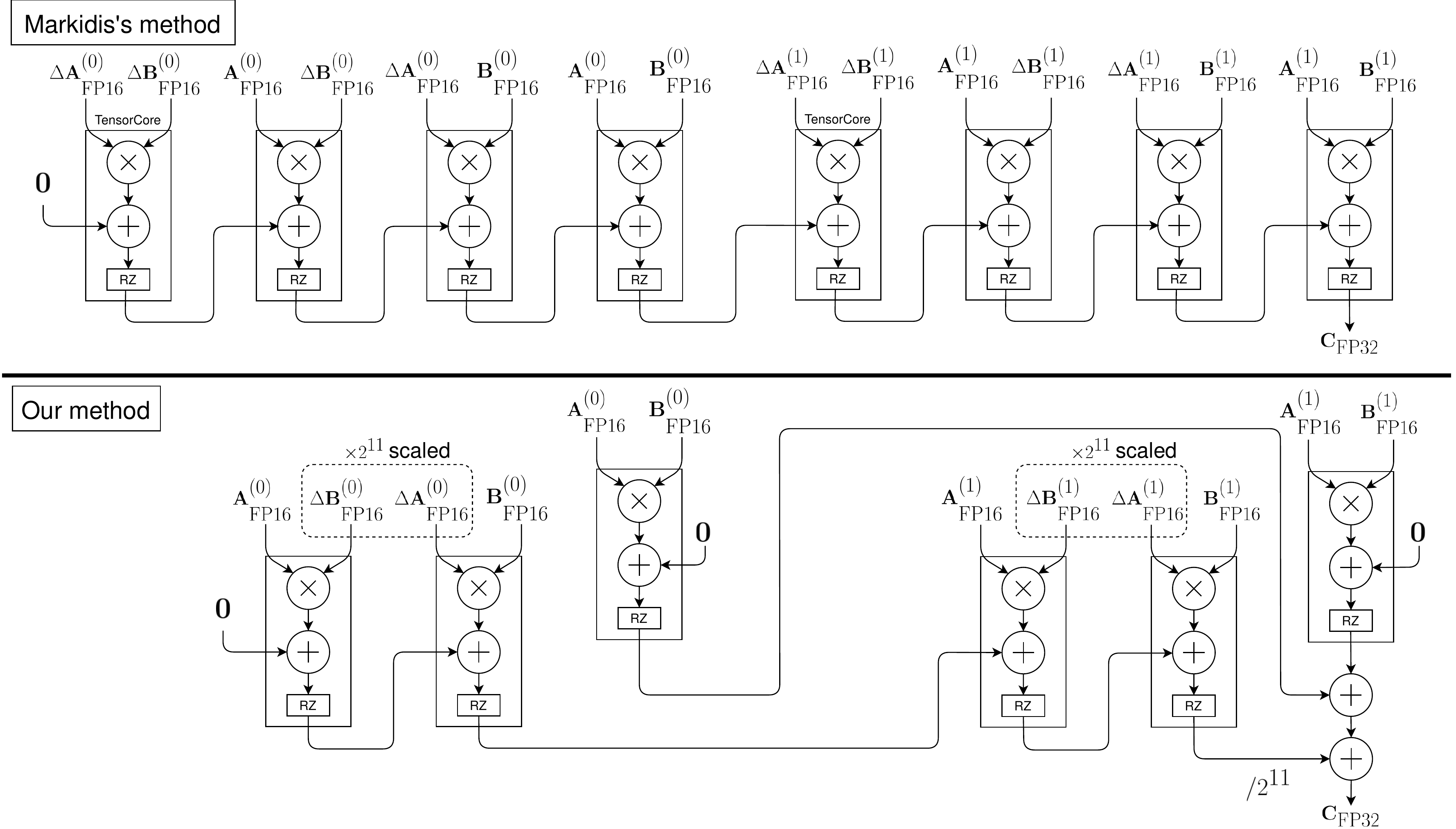}
    \caption{The comparison of Markidis' method (top) and our method (bottom) to carry out single-precision matrix-matrix multiplication $\begin{bmatrix}\mathbf{A}_\mathrm{FP32}^{(0)} & \mathbf{A}_\mathrm{FP32}^{(1)}\end{bmatrix}\begin{bmatrix}\mathbf{B}_\mathrm{FP32}^{(0)} \\ \mathbf{B}_\mathrm{FP32}^{(1)}\end{bmatrix}$. %
The matrices $\mathbf{A}_\mathrm{FP16}^{(\cdot)},\Delta\mathbf{A}_\mathrm{FP16}^{(\cdot)},\mathbf{B}_\mathrm{FP16}^{(\cdot)},\Delta\mathbf{B}_\mathrm{FP16}^{(\cdot)}$ in Markidis' method are calculated by Eqs. (\ref{eq:corr-1})-(\ref{eq:corr-4}), and in our method by Eqs. (\ref{eq:corr-1-1024})-(\ref{eq:corr-4-1024}).%
    }
    \label{fig:mma2}
\end{figure*}

NVIDIA CUTLASS\footnote{\url{https://github.com/NVIDIA/cutlass}} is an open-source CUDA C++ matrix-matrix multiplication template library that has hierarchical memory blocking strategies and computing primitives.
We use CUTLASS version 2.5 as a base implementation and include our techniques for: avoiding RZ after the addition of $\mathbf{C}_\text{F32}$ in Eq. (\ref{eq:matmul-0}), reducing the underflow and gradual underflow probabilities, while ignoring negligible correction terms.
We develop two types of implementations: {\bf cutlass\_halfhalf} and {\bf cutlass\_tf32tf32}.
We use the {\tt mma.sync.aligned.m16n8k8} PTX instruction, which computes matmul-($16, 8, 8$) and addition using FP16 Tensor Cores. We call this implementation cutlass\_halfhalf.
For using TF32 Tensor Cores, we use a PTX instruction which computes the same size of matrix-matrix multiplication and addition as the FP16 and we call this implementation cutlass\_tf32tf32.
We only add the code for the error correction and use the original CUTLASS code for other parts of the computation, such as loading matrix data from global memory to shared memory.
Therefore, our implementation can compute any matrix-matrix multiplication size as long as CUTLASS supports it.

The avoiding of RZ is only applied to $\mathbf{A}_\mathrm{F16} \mathbf{B}_\mathrm{F16}$ in Eq. (\ref{eq:corr-5-1024-reduce}) and it is not applied to $\Delta\mathbf{A}_\mathrm{F16} \mathbf{B}_\mathrm{F16}  + \mathbf{A}_\mathrm{F16} \Delta\mathbf{B}_\mathrm{F16}$.
This allows us to reduce the required registers and computations.
We show an example using FP16 Tensor Cores in Code \ref{lst:cutlass-wmma}.
Furthermore, we show a simple example of a single-precision matrix-matrix multiplication using Markidis' and our methods.
In this figure, we process $\begin{bmatrix}\mathbf{A}_\mathrm{FP32}^{(0)} & \mathbf{A}_\mathrm{FP32}^{(1)}\end{bmatrix}\begin{bmatrix}\mathbf{B}_\mathrm{FP32}^{(0)} \\ \mathbf{B}_\mathrm{FP32}^{(1)}\end{bmatrix}$ using FP16 Tensor Cores.

\begin{lstlisting}[style=CStyle,caption={Pseudocode including our improvements for computing single-precision matrix-matrix multiplication with error correction using FP16 Tensor Cores.},label={lst:cutlass-wmma}]
void matmul(...) {
  fragment frag_a, frag_b, frag_c;
  fragment frag_da, frag_db, frag_dc;
  fragment frag_tmp;
  shared_mem_fp16 smem_a, smem_b;
  shared_mem_fp16 smem_da, smem_db;
  // Initialize accumulator fragments
  fill_fragment(frag_c,0.f);
  fill_fragment(frag_dc,0.f);
  for (k=0;k<K;k++) {
    // Compute eq (19), (21)
    smem_a=toFP16(mem_a[k])
    smem_b=toFP16(mem_b[k])
    load_matrix_sync(frag_a,smem_a);
    load_matrix_sync(frag_b,smem_b);

    // Compute eq (20), (22)
    smem_da=toFP16((mem_a[k]-toFP32(smem_a)*2048)
    smem_db=toFP16((mem_b[k]-toFP32(smem_b)*2048)
    load_matrix_sync(frag_da,smem_da);
    load_matrix_sync(frag_db,smem_db);

    // Compute a part of eq (24)
    mma_sync(frag_dc,frag_da,frag_b,frag_dc);
    mma_sync(frag_dc,frag_a,frag_db,frag_dc);
    // Initialize a temporal accumulator fragment
    fill_fragment(frag_tmp,0.f);
    mma_sync(frag_tmp,frag_a,frag_b,frag_tmp);
    for (i=0;i<frag_c.num_elements;i++) {
      // Accumulation using FP32 SIMT Core
      frag_c.x[i] += frag_tmp.x[i];
    }
  }
  // Compute a part of eq (24)
  for (i=0;i<frag_c.num_elements;i++) {
    frag_c.x[i] += frag_dc.x[i]/2048;
  }
  // Store result to memory
  store_matrix_sync(mem_c,frag_c);
\end{lstlisting}

In our actual implementation, which is more complex than the simple example shown in Code \ref{lst:cutlass-wmma}, we don't store $\mathbf{A}_\text{F16}, \mathbf{B}_\text{F16}$ and $\Delta\mathbf{A}_\text{F16}, \Delta\mathbf{B}_\text{F16}$ explicitly to the shared memory in order to reduce the memory footprint.
Instead, we load $\mathbf{A}_\text{F32}, \mathbf{B}_\text{F32}$ from the shared memory, compute Eq. (\ref{eq:corr-1-1024})-(\ref{eq:corr-4-1024}) on the registers, and store them to the fragments directly.

\subsubsection{Parameter tuning of CUTLASS}

\begin{table*}[t]
\begin{center}
\begin{tabular}{llp{10cm}}
\toprule
Parameter & Values                        & Description                                    \\
\midrule
{\tt bm}, {\tt bn}, {\tt bk} & 16, 32, 64, 128, for respectively & The size of thread block level blocking. Each thread block computes matmul-({\tt bm}, {\tt bn}, {\tt bk}).               \\
{\tt wm}, {\tt wn}, {\tt wk} & 16, 32, 64, 128, for respectively & The size of warp level blocking. Each warp computes matmul-({\tt wm}, {\tt wn}, {\tt wk}). \\
{\tt stages}     & 3, 4                          & The number of software pipelines. \\
\bottomrule
\end{tabular}
\caption{The parameter space for grid search when optimizing the computing performance of CUTLASS.}
\label{tab:wandb-params-space}
\end{center}
\end{table*}
The CUTLASS library has some template parameters to determine the blocking size of each memory layer, the number of software pipeline stages, etc.
We searched the parameters for achieving the highest throughput for each input matrix size using a grid search.
We used Weights\&Biases'\footnote{\url{https://wandb.ai/site}} sweeps for searching the optimal parameters efficiently.
We show the parameter search space in Table \ref{tab:wandb-params-space}.
The total number of all parameter combinations is 3,456 and we filter them with the following set of rules.
\begin{itemize}
    \item At least, one of {\tt wm} $>$ {\tt bm}, {\tt wn} $>$ {\tt bn}, and {\tt wk} $>$ {\tt bk} is satisfied.
    This is because the size of thread-block level blocking must be larger than the warp level.
    \item The size of the shared memory required exceeds its capacity.
    \item The error calculated by Eq. (\ref{eq:eval-matmul}) is larger than the threshold (even if the compilation is successful).
    At this point, we set $0.1$ as the threshold and we checked experimentally that this value holds for all cases.
\end{itemize}
Through this automatic filtering process, we were able to reduce the number of parameter combinations for cutlass\_halfhalf to 202, and for cutlass\_tf32tf to 200.
\section{Experiment details}

We compare the accuracy, throughput, and power consumption of our implementations.
The list of implementations we used are summarized in Table \ref{tab:implementation}.

\begin{table}[]
\center
\begin{tabular}{lcc}
\toprule
Implementation     & TensorCore & Error Correction \\
\midrule
cutlass\_tf32tf32  & TF32-TC    & YES              \\
cutlass\_fp16fp16  & FP16-TC    & YES              \\
cublas\_tf32tc       & TF32-TC    & NO               \\
cublas\_fp16tc       & FP16-TC    & NO               \\
cublas\_simt(FP32) & Not used   & NO       \\
\bottomrule
\end{tabular}
\caption{The list of implementations we use to evaluate our proposed methods. The implementations named cutlass\_XX are our implementation, and cublas\_XX are reference implementations.}
\label{tab:implementation}
\end{table}

\subsection{Accuracy evaluation}
We input single-precision matrices for each implementation and calculate the error following Eq. (\ref{eq:eval-matmul}).
We compute matrix-matrix multiplication $8$ times with different random seeds and average the error of each of them.
The order of addition is changed by the template parameters of CUTLASS, which slightly affects the error.
We use the worst values in the grid search as the error.
We use NVIDIA A100 40GB SXM4 and CUDA version 11.3.

In this section, we evaluate the effect of the exponent range and its pattern.

\subsubsection{Effect of the exponent range of input matrices}
\begin{figure*}
    \centering
    \includegraphics[width=\linewidth]{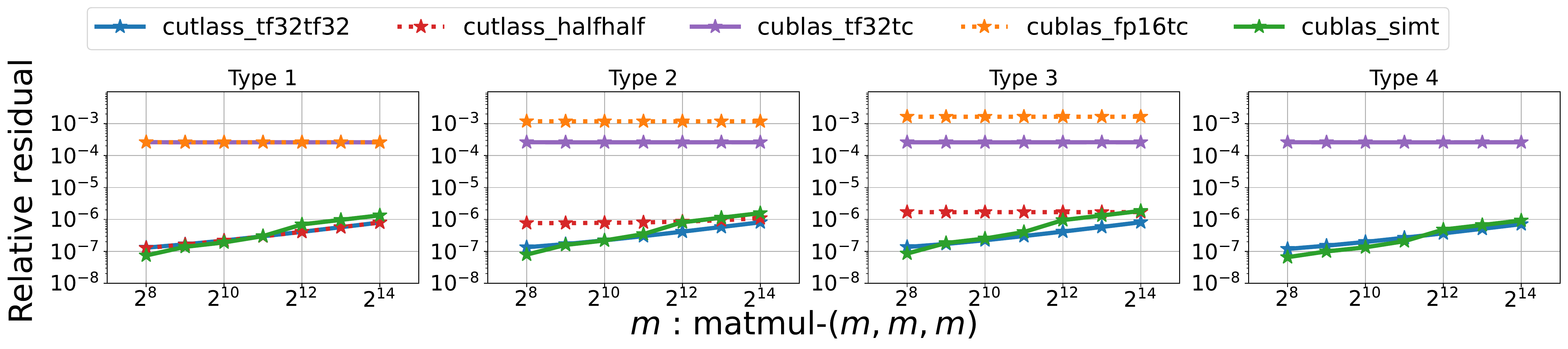}
    \caption{
    The effect of the exponent range on the accuracy of matrix-matrix multiplication $\mathbf{A}_\text{F32} \times \mathbf{B}_\text{F32}$.%
 {\bf Type 1}: All elements in both $\mathbf{A}_\text{F32}$ and $\mathbf{B}_\text{F32}$ are represented with high precision in halfhalf.%
 {\bf Type 2}: All elements in either $\mathbf{A}_\text{F32}$ or $\mathbf{B}_\text{F32}$ are represented with high precision, while the others are lower precision for smaller values in halfhalf.%
 {\bf Type 3}: All elements of both $\mathbf{A}_\text{F32}$ and $\mathbf{B}_\text{F32}$ are lower precision for smaller values in halfhalf.%
 {\bf Type 4}: At least one of $\mathbf{A}_\text{F32}$ or $\mathbf{B}_\text{F32}$ is all zero in halfhalf because they are out-of-range.%
    }
    \label{fig:exp_rand-residual}
\end{figure*}
As we mentioned in the previous section, the exponent range of halfhalf is narrower than FP32 as shown in Figure \ref{fig:halfhalf-precision}.
To evaluate this effect on the matrix-matrix multiplication accuracy, we input the matrices with various exponent ranges and evaluate the accuracy.
We define a single-precision matrix set exp\_rand$(a, b)$ ($a, b \in \mathbb{Z}$) so that the exponent of each element of a matrix in it is generated from a uniform distribution [$a, b$] and the mantissa is generated from a uniform distribution [$0, 2^{23} - 1$].
\begin{eqnarray}
    e &\leftarrow& \mathrm{UniformRandInt}[a, b] \nonumber \\
    m &\leftarrow& \mathrm{UniformRandFP32}[1, 2) \nonumber \\
    s &\leftarrow& \mathrm{UniformRandInt}[0, 1] \nonumber \\
    \label{eq:exp-rand}
    (i, j)\text{-element} &\leftarrow& (2s - 1) \times 2^{e} \times m
\end{eqnarray}

In this evaluation, we use three types of input matrices as follows:
\begin{description}[style=nextline]
\item[exp\_rand($-15, 14$)]
    All elements are in range of $(10^{-5}, 10^{5})$, and represented by our halfhalf with high precision as shown in Figure \ref{fig:halfhalf-precision}.
\item[exp\_rand($-35, -15$)]
    All elements are in range of $(10^{-11}, 10^{-4})$, and represented by our halfhalf with lower precision for smaller values as shown in Figure \ref{fig:halfhalf-precision}.
\item[exp\_rand($-100, -35$)]
    All elements are in range of $(10^{-31}, 10^{-10})$.
    The halfhalf can't represent this range of numbers.
\end{description}

We initialize the input matrices $\mathbf{A}_\text{FP32}$ and $\mathbf{B}_\text{FP32}$ with the above three patterns, respectively, and compute the matrix-matrix multiplication $\mathbf{A}_\text{FP32} \mathbf{B}_\text{FP32}$.
We show the accuracy for the following combinations:
\begin{description}[style=nextline]
\item[Type 1]
    Both $\mathbf{A}_\text{FP32}$ and $\mathbf{B}_\text{FP32}$ are exp\_rand($-15, 14$).
\item[Type 2]
    One of $\mathbf{A}_\text{FP32}$ or $\mathbf{B}_\text{FP32}$ is exp\_rand($-15, 14)$ and the other one is exp\_rand($-100, -35$).
\item[Type 3]
    Both $\mathbf{A}_\text{FP32}$ and $\mathbf{B}_\text{FP32}$ are exp\_rand($-35, -15$).
\item[Type 4]
    At least one of $\mathbf{A}_\text{FP32}$ or $\mathbf{B}_\text{FP32}$ is exp\_rand($-100, -35$).
\end{description}
The outcome of this experiment is shown in Figure \ref{fig:exp_rand-residual}.
We can see that cutlass\_tf32tf32 computes matrix-matrix accuracy with the same accuracy as FP32 SIMT in all cases.
On the other hand, although cutlass\_halfhalf computes in the same accuracy with FP32 SIMT in case Type 1, the loss of accuracy occurs in Type 2 and 3, and cutlass\_halfhalf is not able to perform in case Type 4.
Therefore, if all elements in the matrix have very small exponents, we need to carry out additional scaling before matrix-matrix multiplication is performed.

\subsubsection{Effect of exponent patterns of the input matrices}

\begin{figure*}
    \centering
    \includegraphics[width=\linewidth]{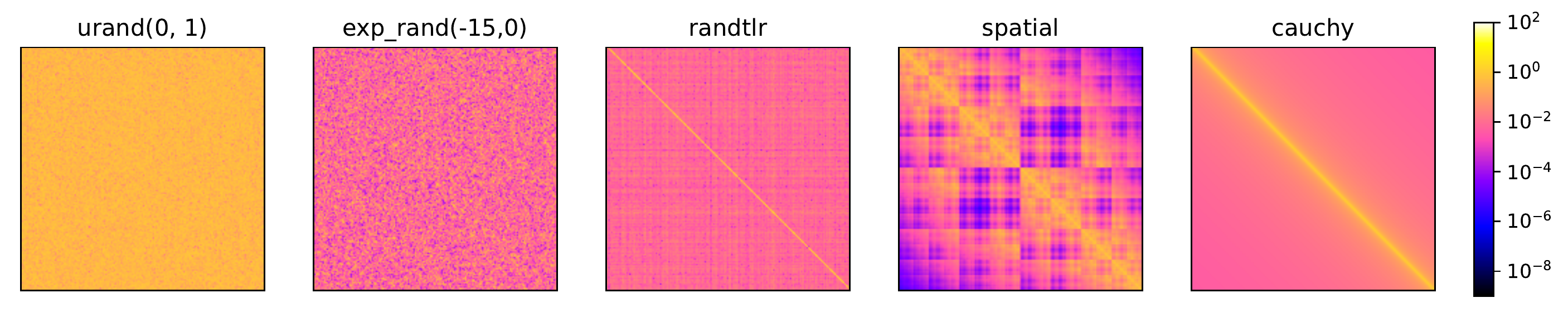}
    \caption{Visualization of the exponent pattern of the input matrices $\mathbf{A}_\text{FP32}$, $\mathbf{B}_\text{FP32}$.%
We use randtlr, spatial, cauchy as $\mathbf{B}_\text{FP32}$, and urand(0,1), exp\_rand(-15, 0) as $\mathbf{A}_\text{FP32}$.}
    \label{fig:stars-h-exp-dist}
\end{figure*}

\begin{figure*}
    \centering
    \includegraphics[width=\linewidth]{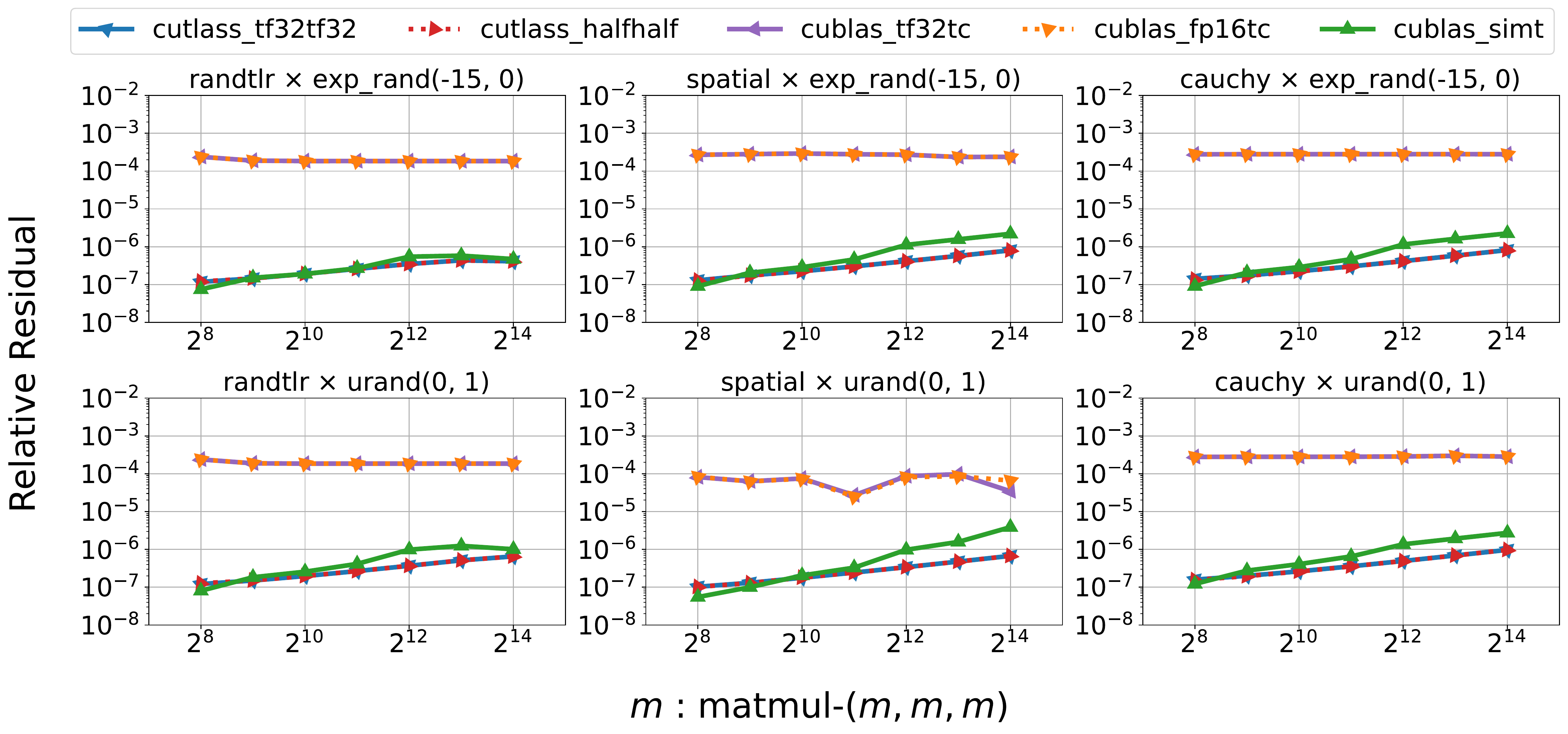}
    \caption{The matrix-matrix multiplication accuracy evaluation using input matrices generated by STARS-H.}
    \label{fig:full-eval}
\end{figure*}

In real-world computations, matrices have various exponent patterns.
STARS-H\footnote{\url{https://github.com/ecrc/stars-h}} was originally developed for the evaluation of hierarchical low-rank approximation methods, and covers various dense matrix patterns in real applications.
We generate three types of input matrix $\mathbf{A}_\text{F32}$ using STARS-H as follows:
\begin{description}[style=nextline]
\item[randtlr]
    Random synthetic TLR (Tile Low-Rank) matrix.
\item[spatial]
    Exponential kernel for spatial statistics.
\item[cauchy]
    Cauchy matrix.
\end{description}
And as input matrix $\mathbf{B}_\text{F32}$, we use two types of input matrix as follows:
\begin{description}[style=nextline]
\item[urand($-1, 1$)]
    Random matrix from a uniform distribution (-1, 1).
\item[exp\_rand($-15, 0$)]
    Random matrix generated by Eq. (\ref{eq:exp-rand}).
\end{description}
We show a sample of exponent patterns of these matrices in Figure \ref{fig:stars-h-exp-dist} and the accuracy of matrix-matrix multiplication $\mathbf{A}_\text{F32}\mathbf{B}_\text{F16}$ in Figure \ref{fig:full-eval}.

Although the accuracy of cutlass\_halfhalf and cutlass\_tf32tf32 are better than cublas\_simt in some matrix sizes, this likely due to the order of addition, and not because of our error correction method.
Thus, we conclude that the accuracy of cutlass\_halfhalf and cutlass\_tf32tf32 are the same as cuBLAS SGEMM, for various patterns of exponent values in the matrix.

\subsection{Performance evaluation}
\label{sec:performance-eval}
\begin{figure*}
    \centering
    \includegraphics[width=\linewidth]{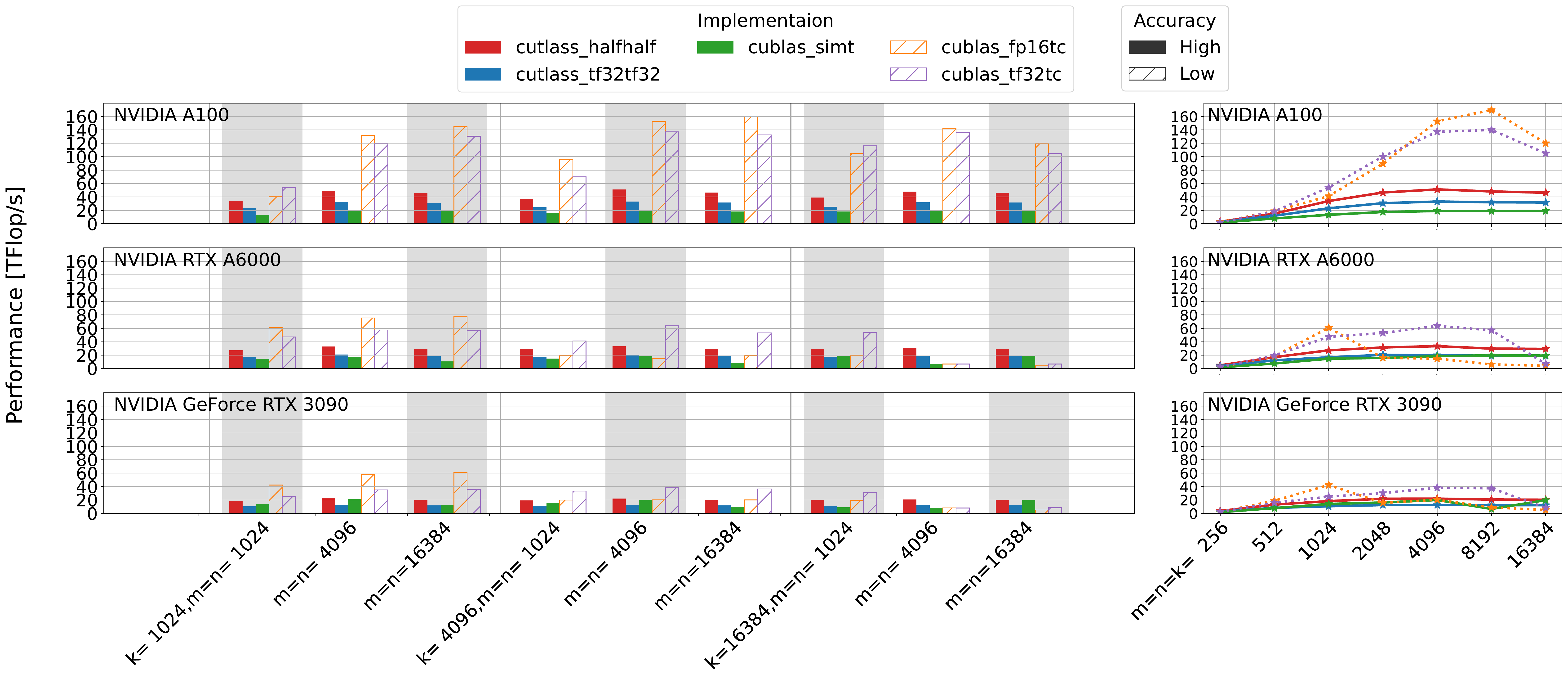}
    \caption{The evaluation of computational throughput on NVIDIA A100, RTX A6000, and GeForce RTX 3090 for matmul-($m, n, k$).%
The solid lines compute single-precision matrix-matrix multiplication in FP32 accuracy (high accuracy) while the dashed lines are for FP16 accuracy (low accuracy).}
    \label{fig:performance-eval}
\end{figure*}

We calculate the Flop/s for matmul-$(m, m, m)$ by dividing $2m^3$ with the computing time $s$ [sec].
The performance on NVIDIA A100, RTX A6000, and GeForce RTX 3090 is shown in Figure \ref{fig:performance-eval}.
The specifications of the host machine: for each GPU are:
\begin{description}[style=nextline]
\item[NVIDIA A100 (40GB, SXM4)] AMD EPYC 7742, 1 TB of DDR4 Memory (3200 MT/s)
\item[NVIDIA RTX A6000] AMD EPYC 7302, 256 GB of DDR4 Memory (3200 MT/s)
\item[NVIDIA GeForce RTX 3090] AMD EPYC 7402, 512 GB of DDR4 Memory (3200 MT/s)
\end{description}

\begin{figure}
    \centering
    \includegraphics[width=0.6\linewidth]{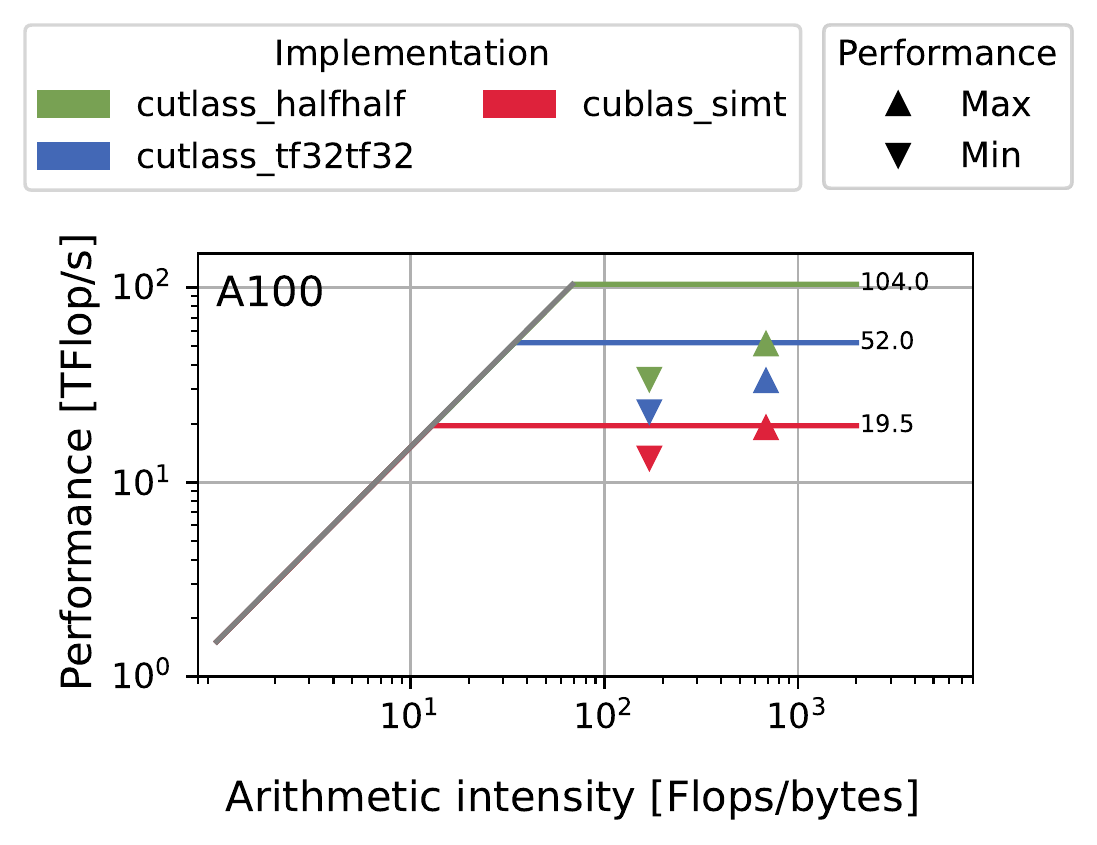}
    \caption{Roofline performance analysis for maximum and minimum execution.%
The theoretical peak performances of cutlass\_fp16fp16 and cutlass\_tf32tf32 are approximated by dividing the peak performance of Tensor Cores by 3.}
    \label{fig:roofline}
\end{figure}

On the NVIDIA A100, cutlass\_halfhalf and cutlass\_tf32tf32 are faster than cublas\_simt for all matrix sizes. Furthermore, they are faster than the theoretical peak performance of FP32.
The performance of cutlass\_halfhalf achieves 51 TFlop/s and cutlass\_tf32tf32 achieves 33 TFlop/s at maximum peak.
The theoretical peak performance of FP16 Tensor Core and TF32 Tensor Core on NVIDIA A100 is 312 TFlop/s and 156 TFlop/s respectively \cite{nvidia_corporation_nvidia_nodate-1}.
Therefore, the theoretical upper limit of our method is $312 / 3 = 104$ TFlop/s for cutlass\_halfhalf and $156 / 3 = 52$ TFlop/s for cutlass\_tf32tf32 since we need three times more computation for the error correction as shown in Eq. (\ref{eq:corr-5-1024-reduce}).
Thus, the ratio against the theoretical peak is 49\% for cutlass\_halfhalf and 63\% for cutlass\_tf32tf32.
For the evaluation on the other GPUs, although cutlass\_halfhalf is faster than cublas\_simt for all matrix sizes, cutlass\_tf32tf32 is slower than cublas\_simt in some cases.
We show the specifications of each GPU in Table \ref{tab:gpu-spec}.
The theoretical peak performance of cutlass\_tf32tf32 on RTX3090, which is $71/3=23.7$ TFlop/s, is lower than FP32, which is $35.58$ TFlop/s.
The GA102 architecture, which RTX 3090 and A6000 are equipped with, can execute FP32 operations on the datapath for integer operations.
And the theoretical peak performance of FP32 shown in Table \ref{tab:gpu-spec} is calculated as a sum of FP32 datapath and integer datapath performance.
Therefore, if the performance of cuBLAS on RTX 3090 is improved by using this feature more, it might be impossible for cutlass\_tf32tf32 to outperform the cuBLAS SGEMM.
Furthermore, we show a roofline performance analysis\cite{williams_roofline_2009} on NVIDIA A100 in Figure \ref{fig:roofline}.
Our implementations do not reach the theoretical peak performance and memory bandwidth.
Therefore, we acknowledge that there is still room for improvement in the implementation.

\begin{table}[]
\center
\begin{tabular}{llll}
\toprule
                 & FP16-TC & TF32-TC  & FP32  \\
                 &  {[}TFlop/s{]} &  {[}TFlop/s{]}  & {[}TFlop/s{]}\\
\midrule
A100       & 312                   & 156                   & 19.5                    \\
RTX A6000  & 309.6                 & 154.8                 & 38.7    \\
RTX 3090   & 142                   & 71                    & 35.58   \\
\midrule
\midrule
                 & Bandwidth & L1 Cache & L2 Cache \\
                 & {[}GB/s{]} & {[}KB/SM{]} & {[}MB{]}\\
\midrule
A100       & 1555 & 164 & 40 \\
RTX A6000  & 768 & 128 & 6 \\
RTX 3090   & 932 & 128 & 6 \\
\bottomrule
\end{tabular}
\caption{The GPU specifications used in our evaluation \cite{nvidia_corporation_nvidia_nodate,nvidia_corporation_nvidia_nodate-1,nvidia_corporation_nvidia_nodate-2}.}
\label{tab:gpu-spec}
\end{table}

\subsection{The power consumption evaluation}
\begin{figure*}
    \centering
    \includegraphics[width=\linewidth]{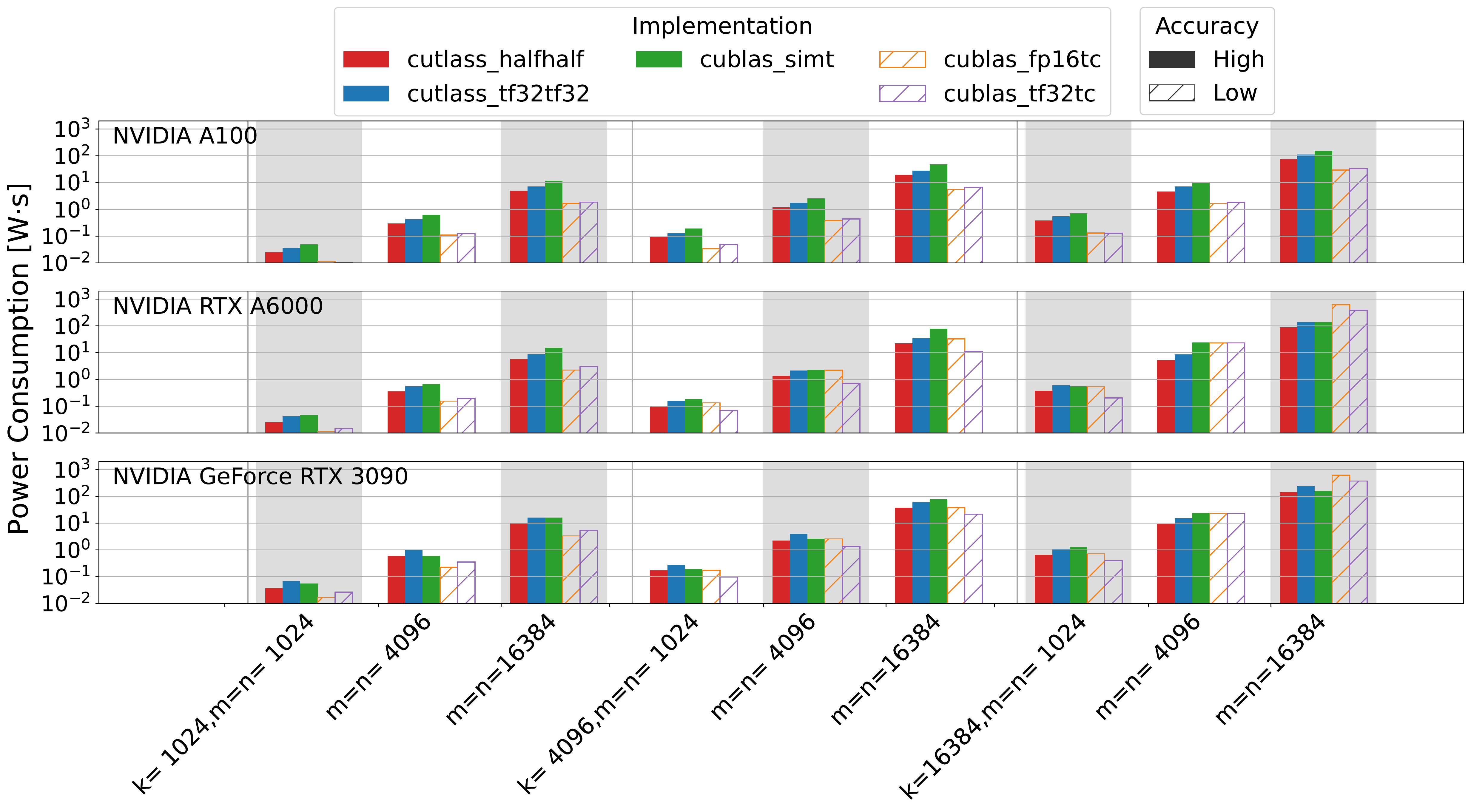}
    \caption{The evaluation of power consumption on NVIDIA A100, RTX A6000, and GeForce RTX 3090 for matmul-($m, n, k$).%
The solid lines show single-precision matrix-matrix multiplication in FP32 accuracy (high accuracy) while the dashed lines are for FP16 accuracy (low accuracy).}
    \label{fig:power}
\end{figure*}

\begin{table*}[]
\begin{center}
\begin{tabular}{cccccc}
\toprule
\multirow{2}{*}{Implementation}    & \multirow{2}{*}{Accuracy}   & \multicolumn{2}{c}{Computing Performance}        & \multicolumn{2}{c}{Power Consumption}                  \\
\cmidrule(lr){3-6}
                                   &                             & A100             & Other GPUs                    & A100                   & Other GPUs                    \\
\midrule
\multirow{2}{*}{cutlass\_tf32tf32} & \multirow{2}{*}{\color{darkGreen}{Same}}       & \color{darkGreen}{Faster}           & \multirow{2}{*}{Case-by-case} & \multirow{2}{*}{\color{darkGreen}{Lower}} & \multirow{2}{*}{Case-by-case} \\
                                   &                             & (Max: 33TFlop/s) &                               &                        &                               \\
\multirow{2}{*}{cutlass\_halfhalf} & \color{darkGreen}{Same}                        & \color{darkGreen}{Faster}           & \multirow{2}{*}{\color{darkGreen}{Faster}}       & \multirow{2}{*}{\color{darkGreen}{Lower}} & \multirow{2}{*}{\color{darkGreen}{Lower}}        \\
                                   & (exponent range is limited) & (Max: 51TFlop/s) &                               &                        &                               \\
\bottomrule
\end{tabular}
\end{center}
\caption{Comparison between our implementations and cuBLAS single-precision matrix-matrix multiplication.}
\label{tab:consclusion}
\end{table*}

We calculate the power consumption per matrix-matrix multiplication using the best CUTLASS template parameters (Figure \ref{fig:power}).
We use NVML (NVIDIA Management Library) to monitor the power consumption every 0.02 sec and aggregate the data\footnote{\url{https://github.com/enp1s0/gpu_monitor}}.
To obtain enough data for the evaluation, we compute a sequence of matrix-matrix multiplication for at least 2 sec.
We evaluate the power consumption on NVIDIA A100, RTX A6000, and GeForce RTX 3090.

Similar to the computational throughput in the previous section, the power consumption of cutlass\_halfhalf and cutlass\_tf32tf32 on NVIDIA A100 is lower than cublas\_simt for all matrices.
The performance-per-power-consumption of cutlass\_halfhalf achieves 121 GFlops/W and cutlass\_tf32tf32 achieves 80.9 GFlops/W at maximum peak while cublas\_simt is 67.0 GFlops/W.
Although the power consumption of cutlass\_halfhalf is also lower than cublas\_simt on the other GPUs for all matrices, cutlass\_tf32tf32 is higher than cublas\_simt for some matrices.
The power consumption and computing time are proportional in many cases.

\subsection{Summary of experiments}
We show the summary of the experiments in Table \ref{tab:consclusion}.
For cutlass\_tf32tf32 on A100, we are able to perform single-precision matrix-matrix multiplication faster than cuBLAS SGEMM, while retaining the exact same accuracy.
For cutlass\_halfhalf, although the exponent range it can handle is limited, it computes single-precision matrix-matrix multiplication faster than cuBLAS with lower power consumption on all three GPUs.

\section{Conclusion}

In this paper, we improve upon the error correction methods proposed by Markidis \textit{et al.} and Feng \textit{et al.} for matrix-matrix multiplication on Tensor Cores.
To the extent of our knowledge, this is the first time the accuracy of GEMM on Tensor Cores has matched the accuracy of cuBLAS SGEMM through error correction while outperforming the FP32 theoretical peak performance.
We use CUTLASS as the base for our implementation, and reach approximately 56\% of the theoretical peak performance on Tensor Cores.

We also perform a thorough analysis of the expected error caused by different rounding modes.
We calculate the expectation of mantissa length kept in our method under an i.i.d. assumption of mantissa bits and prove that the loss in mantissa length is not the main cause of error in the previous work by Markidis \textit{et al.} and Feng \textit{et al.}.
To improve the accuracy, we avoid the RZ rounding performed inside Tensor Cores and reduce the underflow and gradual underflow probabilities.
Furthermore, we reduce the amount of computation during the error correction by ignoring the last term between the differences.

As a result, our implementation computes single-precision matrix-matrix multiplication with the same accuracy as cuBLAS SGEMM while exceeding the throughput of cuBLAS SGEMM, and also exceeding the theoretical peak performance of FP32 on NVIDIA A100.
Our implementation also consumes lower power compared to cuBLAS SGEMM.

\subsection*{Acknowledge}
We would like to thank to Edgar Josafat Martinez Noriega, Hana Hoshino and Sameer Deshmukh for helpful discussions and comments.
This work was partially supported by JSPS KAKENHI JP18H03248, JP21H03447, JP21J14694 and JST CREST JPMJCR19F5.
This work was partially supported by "Joint Usage/Research Center for Interdisciplinary Large-scale Information Infrastructures" in Japan (Project ID: jh210024-NAHI)

\bibliographystyle{plain}
\bibliography{ref}

\end{document}